\documentclass[12pt,letterpaper]{article}
\usepackage{epsfig,rotating,setspace,latexsym,amsmath,epsf,amssymb,amsfonts,bm,theorem,cite,caption,subcaption,enumerate,longtable,accents,algorithm,graphicx,epsf,authblk,epstopdf,url,color,multirow,algpseudocode,mathtools,comment,tabularx,comment,physics,graphicx,caption,subcaption,braket,enumitem,xcolor,algorithm, dsfont,bbm}
\usepackage[short,c2]{optidef}
\usepackage[toc,title]{appendix}
\usepackage{dsfont}

\newtheorem{theorem}{Theorem}

\newtheorem{corollary}{Corollary}
\newtheorem{definition}{Definition}
\newtheorem{remark}{Remark}
\newtheorem{lemma}{Lemma}
\newtheorem{example}{Example}

\newenvironment{Proof}[1]{\medskip\par\noindent{\bf Proof:\,}\,#1}{{\mbox{\,$\blacksquare$}\par}}

\allowdisplaybreaks

\title{Private Information Retrieval With Arbitrary Privacy Requirements: Introduction and Capacity Results\footnote{A part of this work was accepted in IEEE ITW 2026.}}

\author{Mohamed Nomeir \qquad Shreya Meel \qquad Sennur Ulukus \\
\normalsize Department of Electrical and Computer Engineering \\
\normalsize University of Maryland, College Park, MD 20742 \\
\normalsize {\qquad {\it mnomeir@umd.edu} \qquad \it smeel@umd.edu} \qquad {\it ulukus@umd.edu}}

\begin{document}
\date{}
\maketitle
\vspace{-0.8cm}

\begin{abstract}
    In this paper, we introduce the problem of private information retrieval (PIR) under arbitrary privacy requirements, in a graph-based storage system. This formulation is motivated by the server storage limitations, abundance of data (messages) and heterogeneous data privacy requirements. Under the arbitrary privacy requirement, each message has to be retrieved privately from a pre-specified subset of servers, where the subset always includes the servers storing it. Thus, each server is associated with a privacy set, which pre-specifies the message indices that should be privately retrieved from it. This setting is a generalization of the classical PIR setting, where the required message index needs to be kept private from \emph{all} servers, i.e., there, the privacy set of each server comprises all message indices. Our setting is also a bridge between the newly formulated local PIR (LPIR) setting and the classical PIR setting, where in the former, the privacy set is exactly the set of stored message indices. In this paper, we derive general lower and upper bounds on the PIR capacity for general graphs, under certain privacy requirements, that capture the essence of both LPIR and classical PIR. Then, we focus on path and cyclic storage graphs under these and more fine-grained settings, for which we derive capacity results for certain cases, and establish lower and upper bounds for others. Their low degree allows for a more in-depth understanding of the new privacy formulation and admits more privacy requirement settings compared to other simple graphs. Finally, we introduce a new graph structure, the pyramid storage graph, to model server storage. Although this graph has never been investigated in the literature in any PIR context, it enjoys a nice symmetric structure for message storage and replication patterns. In this setting, we provide a capacity lower bound for LPIR. Interestingly, the corresponding scheme is shown to be also valid for two other more stringent privacy settings.
\end{abstract}

\section{Introduction and Motivation}
The initial formulation of the private information retrieval (PIR) problem \cite{chor} considers $N$ fully-replicated servers, where all $K$ messages are stored at all servers. The goal is to design a scheme to retrieve one out of $K$ messages without letting any individual server know the retrieved message index. The metric is the communication efficiency, i.e., the rate, which is the ratio of the average length of the required message to the average total number of downloaded symbols. The highest achievable rate is called the capacity of PIR. The capacity of the fully-replicated PIR is derived in \cite{c_pir}. Subsequently, several variants of the fully-replicated PIR under various threat models are analyzed in \cite{nan_eaves, banawan_eaves, byzantine_tpir, semantic_pir, semantic_tpir, nomeir_asymp_bspir, ulukusPIRLC, csa, C_SETPIR, skoglund_mds_spir, asymmetric}, and the corresponding quantum versions are analyzed recently in \cite{our_quantum_first, lu_jafar_byzantine_quantum, our_symm_byz_quantum_journal,qpir}. As a major departure from the fully-replicated setting, \cite{banawan_pir_mdscoded} considers the case where the messages are encoded using maximum distance separable (MDS) codes for each message individually and stored in the $N$ servers. For an $(M,N)$ MDS code, this significantly reduces server storage (by a factor of $M$) at the cost of reduced PIR capacity. Even though the server storage is reduced, a portion of every message is still stored in every server; therefore, this may be viewed as a version (coded) of fully-replicated setting. Further, \cite{mds_breaking} considers the case of jointly encoded storage of messages, where the messages are mixed during encoding. In another line of research, weakly PIR  \cite{LKRAY21_TIT,QZTL22} aims to relax the privacy requirement. In weakly PIR, a pre-defined privacy leakage budget is allowed, which provides a higher rate by leveraging the leakage budget. The weakly PIR provides a uniform leakage budget to all messages in the system. 

As a first real step in moving away from full-replication,  \cite{graphbased_pir} introduces the idea of graph PIR (GPIR), where each message is replicated in a subset of servers, i.e., each server holds a subset of messages. This is equivalent to a hyper-graph, where the nodes represent the servers and the hyper-edges represent the messages; each message represented by a hyper-edge is stored at the servers that the hyper-edge is incident on. Reference \cite{karim_graph} considers a different modeling convention, where the messages are the nodes and the servers are the edges in a graph representation. These works are followed by extended settings for GPIR for simple- and multi-graphs \cite{cyclic_optimal_length, sadeh_graphs, jafar_asymptotic_graph, shreya_graph_3, shreya_graph_5, gePIR, krishnan_graph, shreya_common_randomness_gspir}. In all the aforementioned formulations, one requirement remained constant: the required message index must be kept \emph{private} against \emph{all} servers. This requirement becomes too restrictive when the graph contains a large number of nodes $N$ and when the message replication is limited. This is evident from the available capacity results, where the capacity decays as fast as $\Theta(1/N)$. This motivates the need for a new privacy model to accommodate data abundance, storage limitation, and privacy heterogeneity.

Recently, the concept of \emph{local privacy} and the accompanying local PIR (LPIR) problem have been introduced in \cite{local_pir_conf,local_pir_jour}. In LPIR, the required message index must be kept \emph{private} only from the servers that \emph{store} the required message. For LPIR, \cite{local_pir_conf,local_pir_jour} show a significant increase in the retrieval rate compared to the conventional GPIR. In the present paper, we define a model that accommodates more general privacy requirements, and encompasses LPIR and GPIR as special cases. The key observation is that, since data is heterogeneous, the privacy budget can also be heterogeneous; for example, retrieving a specific patient's medical record might require higher privacy than retrieving a movie. Motivated by such real-world settings, we define the problem of \emph{PIR with arbitrary privacy requirements}, where each message has its own privacy requirement based on its sensitivity. Thus, to fully define a setting in our framework, one needs both the storage pattern and the privacy pattern; that is, even for a fixed storage graph, we may have multiple required privacy patterns. Finally, we note that there are two important differences between our formulation and the weakly PIR formulation: first, in our case, the privacy budget is message-dependent, i.e., each message has its own privacy requirement; second, the requirement is more tangible than assigning an overall privacy-budget value as in the weakly PIR setting, in the sense that, we precisely know which messages must be kept private for each server when being queried.

In this paper, we introduce PIR with arbitrary privacy requirements as a generalization of GPIR and LPIR, and establish rate bounds and exact capacity results. First, we provide a general result for any graph, for any privacy requirement, where we optimize over all GPIR schemes, tuned to meet the specific privacy setting. For our subsequent results, we introduce three special settings: \emph{extremely private messages}, where a subset of messages is kept private from all servers when retrieving, \emph{oblivious servers}, where a subset of servers have all the messages in their privacy sets, and \emph{modified edge-servers privacy}, where the privacy sets of the leaves are the same as their parent nodes in graphs that are trees. We present a capacity upper bound for the first setting, and lower bounds for the second and third settings. To derive sharper capacity bounds, we focus on two simple graphs: path and cyclic. Specifically, for the path graph, we identify the exact capacity for the extremely private messages setting with odd $N$, and the modified edge-servers privacy for all $N$. For the cyclic graph, we focus on $N\geq 4$, since for $N=3$, the capacity for any arbitrary privacy setting can be shown to be $\frac{1}{2}$, due to matching GPIR and LPIR capacities. Furthermore, we derive the exact capacity for the extremely private messages setting, and for certain subsets of oblivious servers. We also define a new privacy setting where the server keeps the $i$th neighboring server's storage in its privacy set, and derive the corresponding capacity bounds. For both path and cyclic graphs, we also derive capacity bounds for the one- and two-sided $h$-neighbor privacy settings, where the privacy set of each server consists of the indices of messages in the storage of up to its $h$th neighbor. In particular, the two-sided $h$-neighbor privacy setting renders a transition from LPIR to GPIR setting. Finally, we introduce the \emph{pyramid} storage graph, which is a non-uniform hypergraph with a symmetrical, pyramid-like storage architecture. This scenario may arise in practical distributed storage systems, where the cost of replicating each message is different. Alternatively, this can reflect a scenario where all messages may be fully-replicated initially, but some of the servers may have out-dated or erased versions of some messages, making them unsuitable for retrieval. For this graph, we propose an achievable LPIR scheme, which is also a valid scheme for two other settings of the modified edge-servers privacy.

The rest of the paper is organized as follows. Section \ref{sec_formulation} provides the necessary definitions and the problem formulation. Section \ref{sec_examples} provides examples for different privacy settings to motivate and introduce the new framework. Section \ref{sec_main_results} gives the main results for general graphs. Section \ref{sec_results_path} presents specific results for path storage graphs and Section \ref{sec_results_cyclic} presents specific results for cyclic storage graphs. Section \ref{sec_pyramid} introduces and analyzes the pyramid storage graphs. Section \ref{sec_conc} concludes the paper and provides directions for future work.

\emph{Notations:} We use calligraphic font to denote sets. $[a:b]$ denotes the set of integers $\{a,a+1, a+2, \ldots, b\}$ and $[a]$ denotes the set $\{1,2,\ldots, a\}$. In addition, for the family of sets $\mathcal{L}_i$, $i \in \mathcal{I}$, if $\mathcal{Z} \subseteq \mathcal{I}$, then $\mathcal{L}_{\mathcal{Z}} = \cup_{i \in \mathcal{Z}} \mathcal{L}_i$. We use $C_{GPIR}$ to denote the usual PIR capacity over a graph; $C_{LPIR}$ to denote the capacity of LPIR; and $C$ without any subscripts to denote the capacity of our formulation. For the specific graph families, we use $\mathbf{P}_N$ and $\mathbf{C}_N$ to denote the path graph and the cyclic graph with $N$ nodes, respectively.

\section{Problem Formulation}\label{sec_formulation}
\subsection{Preliminaries}

\begin{definition}[Privacy set]
The privacy set $\mathcal{P}_n$ of server $n$ is the set of message indices whose privacy should be maintained at server $n$. 
\end{definition}

\begin{remark}
    Let $\mathcal{I}_n$ denote the set of indices of messages stored in the $n$th server. Then, $\mathcal{I}_n \subseteq \mathcal{P}_n\subseteq [K]$, for all $n\in [N]$. The case of $\mathcal{P}_n =\mathcal{I}_n$ is the setting of local PIR (LPIR), while $\mathcal{P}_n=[K]$ is the setting of classical PIR over graphs (GPIR).
\end{remark}

\begin{definition}[Extremely private messages]
    A message $k \in [K]$ is called an extremely private message if it is contained in the privacy sets of all the servers, i.e., $k \in \mathcal{P}_n$, for all $n \in [N]$. The set of all extremely private messages is $\mathcal{E}$, i.e., $\mathcal{E}= \{k \in [K]: k \in \mathcal{P}_n, n \in [N]\}$.  
\end{definition}

\begin{definition}[Oblivious server]
    A server $n \in [N]$ is called an oblivious server if its privacy set contains all the messages, i.e., $\mathcal{P}_n = [K]$. The set of all oblivious servers is $\mathcal{O}$, i.e., $\mathcal{O} = \{ n\in [N]: \mathcal{P}_n = [K]\}$.
\end{definition}

\begin{definition}[Useful side information]\label{def_side_info}
    Given any GPIR or LPIR scheme $\Pi$, the queries sent to any server can be separated into three classes: queries sent to retrieve symbols for the required message index, i.e., the queries containing the required message index $\theta$, the queries sent to be used in decoding the required message symbols (interference alignment), and the queries sent to ensure privacy. Neither the second nor the third kind of queries contain the required message index. The answers received in response to the second kind of queries are called the useful side information.
\end{definition}

\begin{remark}
    There can be an overlap between the second and third kind of queries in Definition \ref{def_side_info}, in which case, we consider them as the second kind.
\end{remark}

\subsection{Problem Statement}
We consider a  system that consists of $N$ non-colluding servers and $K$ independent messages, where the $i$th message has $L_i$ symbols generated uniformly and independently at random, 
\begin{align}
    H(W_{[K]}) = \sum_{i=1}^K H(W_i) = \sum_{i=1}^K L_i,
\end{align}
where $W_i$ represents the $i$th message, and $[K]$ represents the set $\{1,\ldots,K\}$. As in the case of classical PIR, the user does not have any side information about the message contents, thus, the messages are independent of the queries sent, 
\begin{align}\label{2}
    I(W_{[K]};\mathcal{Q}) = 0,
\end{align}
where $\mathcal{Q} = \{Q_n^{[k]}: n \in [N], k \in [K]\}$ and $Q_n^{[k]}$ is the query sent to the $n$th server when retrieving the $k$th message, including the potential null query $Q_n^{[k]} = \emptyset$. The user transmits the query $Q_n^{[k]}$ such that the required message index remains hidden from the intended servers,
\begin{align}
    I(\theta; Q_n^{[\theta]}| \theta \in \mathcal{P}_n) = 0, \quad n \in [N],
\end{align}
where $\mathcal{P}_n$ is the privacy set. We note that, if a server is not queried, it does not know that the scheme is initiated. However, since the scheme is globally known, when a query is transmitted to a server, it knows whether a message in its privacy set is being queried or not. The answer generated from server $n$ is a function of its individual storage $\mathcal{W}_n$ and the received query, i.e., for all $n\in [N]$, $k\in [K]$,
\begin{align}\label{4}
    H(A_n^{[k]}| ~ \mathcal{W}_n, ~Q_n^{[k]})=0. 
\end{align}
Once the user receives the answers from all servers, the required message must be decodable,
\begin{align}
    \label{decodability}H(W_{\theta}|A_{[N]}^{[\theta]},\mathcal{Q}) = 0.
\end{align}

The rate is defined as in the semantic PIR setting since the messages may be of unequal lengths and the downloaded number of symbols for each message may be different, 
\begin{align}
    R_{\Pi} = \frac{\mathbb
    {E}\left[L\right]}{\mathbb
    {E}\left[D\right]}=\frac{\sum_{i=1}^K L_i}{\sum_{i=1}^K D_i},
\end{align}
where $\Pi$ is a scheme satisfying \eqref{2}-\eqref{decodability} for a given privacy requirement. Since all messages are equally likely, the rate reduces to the second equality, where $L_i$ and $D_i = \sum_{n=1}^N H(A_n^{[i]})$ are the $i$th message length and the number of downloaded symbols when retrieving the $i$th message, respectively. The capacity $C$ is the supremum of the rate over all such $\Pi$.

\section{Motivating Examples}\label{sec_examples}
\begin{example}[Cyclic graph with $1$st neighbor privacy]\label{ex_cycle}
    Let $N=5$, and the storage be given as $\mathcal{W}_1 = \{W_1,W_5\}$, $\mathcal{W}_2 =\{W_1,W_2\}$, $\mathcal{W}_3 = \{W_2,W_3\}$, $\mathcal{W}_4 = \{W_3,W_4\}$, $\mathcal{W}_5 = \{W_4,W_5\}$. The privacy sets are given by
    \begin{align}
        \mathcal{P}_1  = \{1,2,5\}, ~ \mathcal{P}_2  = \{1,2,3\}, ~
        \mathcal{P}_3  = \{2,3,4\}, ~
        \mathcal{P}_4  = \{3,4,5\}, ~ \mathcal{P}_5  = \{1,4,5\}.
    \end{align}
    The message lengths are chosen as $L=2$. Let $W_1=(a_1, a_2)$, $W_2=(b_1,b_2)$, $W_3=(c_1,c_2)$, $W_4=(d_1,d_2)$ and $W_5=(e_1,e_2)$ be message symbols after being uniformly and independently permuted by the user. Table \ref{table_example_cyclic} shows the retrieval scheme for each message index. The rate is $R = \frac{2}{5}$, which is greater than the capacity of the cyclic graph $C_{PIR}(\mathbf{C}_5) =\frac{1}{3}$ \cite{karim_graph}. The rate is lower than the capacity of local PIR $C_{LPIR}(\mathbf{C}_5) =\frac{1}{2}$ due to the locality property \cite{local_pir_jour}.

    \begin{table}[h]
        \centering
        \begin{tabular}{|c|c|c|c|c|c|}
        \hline
             & server 1 & server 2 & server 3  & server 4  & server 5 \\
             \hline
             $\theta = 1$&  $a_1+e_1$& $a_2+b_1$ &$b_1$  &$d_1$  & $d_1+e_1$ \\
             \hline
             $\theta = 2$ &$a_1+e_1$  &$b_1+a_1$  & $b_2+c_1$ & $c_1$ &$e_1$ \\
             \hline
             $\theta = 3$ & $a_1$ & $b_1+a_1$  & $b_1+c_1$ & $c_2+d_1$ & $d_1$\\
             \hline
             $\theta = 4$ & $e_1$ & $b_1$ & $b_1+c_1$ & $c_1+d_1$ & $d_2+e_1$\\
             \hline
             $\theta = 5$ & $a_1+e_1$ & $a_1$ &  $c_1$& $d_1+c_1$ & $d_1+e_2$\\
             \hline
        \end{tabular}
        \caption{Retrieval scheme for Example~\ref{ex_cycle}.}
        \label{table_example_cyclic}
    \end{table}
\end{example}

\begin{example}[Path graph with modified edge-servers privacy]\label{ex_path}
    Let $N=4$ servers, then the storage is given as $\mathcal{W}_1  = \{W_1\}$, $\mathcal{W}_2  = \{W_1,W_2\}$, $\mathcal{W}_3 = \{W_2,W_3\}$, $\mathcal{W}_4  = \{W_3\}$. The privacy sets are given by
    \begin{align}
        \mathcal{P}_1 & = \{1,2\}, ~ \mathcal{P}_2  = \{1,2\}, ~ \mathcal{P}_3  = \{2,3\}, ~ \mathcal{P}_4 = \{2,3\}.
    \end{align}
    The message lengths are chosen as $L=2$. Table \ref{table_exmple_path} shows the retrieval scheme. The rate is $R = \frac{3}{5}$, which is greater than the capacity of the path graph $C_{PIR}(\mathbf{P}_4) = \frac{1}{2}$\cite{shreya_graph_3}. We show in Theorem~\ref{thm:converse_priv_set1_path} that $\frac{3}{5}$ is indeed the capacity of this setting.
 
    \begin{table}[h]
    \centering
    \begin{tabular}{|c|c|c|c|c|}
    \hline
          &  server 1 & server 2  & server 3 & server 4\\
          \hline
          $\theta = 1$& $a_1$ & $a_2+b_1$ & $b_1$& \\
          \hline
          $\theta = 2$ & $a_1$ & $b_1+a_1$ & $b_2+c_1$& $c_1$\\
          \hline
          $\theta = 3$ &  & $b_1$ & $b_1+c_1$ & $c_2$ \\
          \hline
     \end{tabular}
     \caption{Retrieval scheme for Example~\ref{ex_path}.}
     \label{table_exmple_path}
    \end{table}
\end{example}

\begin{example}[$\mathbf{C}_5$ with two oblivious servers]\label{ex_obl}
    Given the cyclic storage graph with $N=5$ as in Example \ref{ex_cycle} with storage $\mathcal{W}_1 = \{W_1,W_5\}$, $\mathcal{W}_2 =\{W_1,W_2\}$, $\mathcal{W}_3 = \{W_2,W_3\}$, $\mathcal{W}_4 = \{W_3,W_4\}$, $\mathcal{W}_5 = \{W_4,W_5\}$, and let the first and the third servers be oblivious and the remaining servers follow local privacy, i.e., the privacy sets are given by
    \begin{align}
        \mathcal{P}_1 = \{1, 2, 3, 4, 5\}, ~ \mathcal{P}_2 = \{1,2\}, ~ \mathcal{P}_3 = \{1, 2, 3, 4, 5\}, ~ \mathcal{P}_4 = \{3,4\}, ~ \mathcal{P}_5 = \{4,5\}.
    \end{align}
    The message lengths are chosen as $L_1=L_2=L_3=L_5=1$ and $L_4=2$. For notational consistency, we define $W_1 = a_1$, $W_2=b_1$, $W_3 = c_1$, $W_4 = (d_1,d_2)$ after permuting, and $W_5=e_1$. Table \ref{tb_ex_oblivious} shows the optimal retrieval scheme for this setting. The rate is $R=\frac{1}{2}$, which matches the capacity for this setting.
    
    \begin{table}[h]
        \centering
        \begin{tabular}{|c|c|c|c|c|c|}
        \hline
             & server 1 & server 2 & server 3 & server 4  & server 5\\
             \hline
             $\theta=1$&  & $a_1,b_1$ &  &  & \\
             \hline
             $\theta=2$&  & $a_1,b_1$ &  &  & \\
             \hline
             $\theta=3$&  &  &  & $c_1,d_1$ & \\
             \hline
             $\theta=4$&  &  &  &$c_1,d_1$  &$d_2,e_1$ \\
             \hline
             $\theta=5$&  &  &  &  & $d_2,e_1$\\
            \hline
        \end{tabular}
        \caption{Optimal retrieval scheme for Example~\ref{ex_obl}.}
        \label{tb_ex_oblivious}
    \end{table}   
\end{example}

\begin{example}
    [$\mathbf{C}_3$ with one extremely private message]\label{ex_c_3_extreme_privacy}
    Consider $\mathbf{C}_3$ with $W_1$ as an extremely private message. The storage is $\mathcal{W}_1 = \{W_1,W_3\}$, $\mathcal{W}_2  = \{W_1,W_2\}$, $\mathcal{W}_3  = \{W_2,W_3\}$. Since the first message index is extremely private, the privacy sets are given by
    \begin{align}
        \mathcal{P}_1 = \{1,3\}, ~ \mathcal{P}_2  = \{1,2\}, ~ \mathcal{P}_3  = \{1,2,3\}.
    \end{align}
    We choose the message length as $L=12$ and the symbols of each message after choosing a permutation uniformly at random are given by $W_1 = a_{[12]}$, $W_2 = b_{[12]}$, and $W_3 = c_{[12]}$. The optimal retrieval scheme is given in Table \ref{tab_ex_c_3_extreme_privacy}. The rate of the scheme is $R=\frac{1}{2}$, which is the capacity for this setting. 

    \begin{table}[h!]
    \centering
    \begin{tabular}{|p{1cm}|p{2.8cm}|p{2.8cm}|p{2.8cm}|}
    \hline
         & server 1 & server 2 & server 3\\
         \hline
         $\theta = 1$& $a_1,a_2,c_1,c_2$, $a_3+c_3$, $a_4+c_4$, $a_5+c_5$, $a_6+c_6$ & $a_7,a_8, b_1,b_2$, $a_9+b_3$, $a_{10}+b_4$, $a_{11}+b_5$, $a_{12}+b_6$ & $b_3,b_4, c_3,c_4$, $b_1+c_5$, $b_2+c_6$, $b_5+c_1$, $b_6+c_2$\\
         \hline
         $\theta=2$& $a_3,a_4, a_5,a_6$, $c_3,c_4$, $c_5,c_6$ & $a_1,a_2, b_1,b_2$, $a_3+b_3$, $a_{4}+b_4$, $a_{5}+b_5$, $a_{6}+b_6$ &  $b_7,b_8, c_1,c_2$, $b_9+c_3$, $b_{10}+c_4$, $b_{11}+c_5$, $b_{12}+c_6$\\
         \hline
         $\theta=3$ & $a_1,a_2, c_7,c_8$, $a_3+c_{9}$, $a_4+c_{10}$, $a_5+c_{11}$, $a_6+c_{12}$ & $a_3,a_4, a_5,a_6$, $b_3,b_4$, $b_5,b_6$ &$b_1,b_2, c_1,c_2$, $b_3+c_3$, $b_{4}+c_4$, $b_{5}+c_5$, $b_{6}+c_6$ \\
         \hline
    \end{tabular}
    \caption{Optimal retrieval scheme for  Example~\ref{ex_c_3_extreme_privacy}.}
    \label{tab_ex_c_3_extreme_privacy}
\end{table}
\end{example}

\section{Results on General Graphs}\label{sec_main_results}
We start by the following lemma. Although it is quite intuitive, it shows how the local PIR and GPIR are connected through the setting of arbitrary private sets PIR.

\begin{lemma}\label{basic_ub_lb_lemma}
    Given a storage hyper-graph $G=([N],\mathcal{W})$, where $\mathcal{W}$ is the set of messages (hyper-edges) with arbitrary private sets $\mathcal{P}_n$, $n \in [N]$, then the capacity of this setting is upper bounded by the local PIR capacity on $G$ and lower bounded by the GPIR capacity over the graph $G$, i.e.,
    \begin{align}
        C_{GPIR}(G) \leq C \leq C_{LPIR}(G).
    \end{align}
\end{lemma}

The proof of Lemma \ref{basic_ub_lb_lemma} is straightforward as $\mathcal{I}_n \subseteq \mathcal{P}_n \subseteq [K]$, i.e., increasing the privacy constraint cannot increase the capacity.

\begin{theorem}\label{thm_relat_gpir}
    Let $\Pi_{PIR}$ be a scheme for GPIR with $K$ messages and $L_i$, $i \in [K]$, message lengths and $D_{n}$, $n \in [N]$, symbols downloaded from the $n$th server. In addition, let $D'_{n,k}$ be the amount of useful side information downloaded from the $n$th server when retrieving the message $\theta = k$. Then, the capacity of PIR with an arbitrary privacy requirement given by $\mathcal{P}_n$, $n \in [N]$, is lower bounded by
    \begin{align}
        C \geq  \max_{\Pi_{PIR}} \ \frac{\sum_{i=1}^KL_i}{\sum_k\left(\sum_n \mathds{1}\{k \in \mathcal{P}_n\} \cdot D_n + \sum_n \mathds{1}\{k \notin \mathcal{P}_n\} \cdot D'_{n,k}\right)}
    \end{align}
\end{theorem}

\begin{Proof}
The proof stems from the fact that any GPIR scheme can be modified to the setting with arbitrary privacy sets $\mathcal{P}_n$, $n \in [N]$, since the privacy requirements are stricter in GPIR. Without loss of generality, let the required index be $\theta = k$, and let $\mathcal{N}_k$ be the set of servers that have $k$ in their privacy set, i.e.,
\begin{align}
    \mathcal{N}_k=\{n \in [N]: k \in \mathcal{P}_n\}.
\end{align}
Since $\mathcal{P}_n \subseteq [K]$, we apply the GPIR scheme $\Pi$ on the servers in $\mathcal{N}_k$, thus downloading $D_n$ for all $n \in \mathcal{N}_k$. On the other hand, we download only the clear side information from the servers $[N]\setminus \mathcal{N}_k$ that can cancel the interference terms from the servers in $\mathcal{N}_k$, to decode $W_k$. This amounts to $D'_{n,k}$, for $n \in [N] \setminus \mathcal{N}_k$. 

Now, since $\Pi$ is symmetric for all messages, by virtue of it being a PIR scheme, for any server with any arbitrary message in its privacy set, it will receive the same queries that provide privacy in the GPIR setting. On the other hand, if it does not have the retrieved message index in its privacy set, we download only the specific required side information needed to decode the required message by canceling the interference.
\end{Proof}

\begin{theorem}\label{general_extreme_privacy}
    Let the privacy sets be defined based on the extremely private messages in $\mathcal{E} \subseteq [K]$ and local privacy, i.e., $\mathcal{P}_n = \mathcal{I}_n \cup \mathcal{E}$, for all $n\in [N]$. In addition, let $D_k$ be the number of downloaded symbols for retrieving $k \in \mathcal{E}$ using the optimal GPIR scheme $\Pi_1$ and let $D'_k$ be the number of downloaded symbols for the optimal local PIR scheme when retrieving $ k \in [K] \setminus \mathcal{E}$ using the optimal local PIR scheme $\Pi_2$, then the capacity for this setting is upper-bounded by
    \begin{align}
        C \leq \frac{\sum_{i \in [K]}L_i}{\sum_{k \in \mathcal{E}}D_k +\sum_{k \in [K] \setminus \mathcal{E}}D'_k},
    \end{align}
    where $L_i$ is the length of $i$th message in the three schemes, i.e, LPIR, GPIR and arbitrary privacy sets PIR.
\end{theorem}

\begin{Proof}
Consider a feasible scheme for this privacy setting that downloads $D_k''$ symbols when retrieving the $k$th message. Then, the total number of download symbols for all messages can be lower bounded as
\begin{align}
\sum_{k \in [K]}D''_k &= \sum_{k \in \mathcal{E}}D''_k+\sum_{k \in [K]\setminus \mathcal{E}}D''_k\\
& \geq \sum_{k \in \mathcal{E}}D_k + \sum_{k \in [K]\setminus \mathcal{E}}D'_k,\label{interence_ub}
\end{align}
where \eqref{interence_ub} is due to the interference upper bounds for each message index and the optimality of the individual schemes. Then,
\begin{align}
    R &= \frac{\sum_{i \in [K]}L_i}{\sum_{i \in [K]}D''_i}\\
    & \leq  \frac{\sum_{i \in [K]}L_i}{\sum_{k \in \mathcal{E}}D_k + \sum_{k \in [K]\setminus \mathcal{E}}D'_k},
\end{align}
which completes the proof.
\end{Proof}

\begin{theorem}\label{general_oblivious}
    For a simple graph $G=([N],[K])$ with $\mathcal{O} \subset [N]$ as the set of oblivious servers, let $D_i$ be the number of downloaded symbols for retrieving the $i$th message, $i\in \mathcal{K}$, in the LPIR scheme, on the subgraph $G'=([N]\setminus \mathcal{O},\mathcal{K})$, where $\mathcal{K} = \cup_{n,m\in [N]\setminus \mathcal{O}} \mathcal{I}_n\cap \mathcal{I}_m$. Further, let $D_{i,n}$ be the corresponding number of symbols downloaded from the $n$th server. If the vertices $[N]\setminus \mathcal{O}$ are a vertex cover for $G$, and letting $\mathcal{K}'=[K]\setminus \mathcal{K}$, the capacity is lower bounded by
    \begin{align} 
        C \geq \frac{\sum_{i}L_i}{\Gamma},
    \end{align}
    where
    \begin{align}
        \Gamma = &\sum_{i\in \mathcal{K}}\left(D_i+ \sum_{n\in [N]\setminus \mathcal{O}:D_{i,n}>0}\sum_{k\in \mathcal{K'}\cap \mathcal{I}_n} L_k\right) \nonumber\\
        &+\sum_{i \in \mathcal{K}'}\sum_{ n\in [N]\setminus \mathcal{O}: ~\exists \ell \in \mathcal{K}\text{ with }D_{\ell,n}>0, \{i,\ell\}\subseteq \mathcal{I}_n}\left( \sum_{j\in \mathcal{K}'\cap \mathcal{I}_n} L_j + D_{\ell,n}\right).
    \end{align}
\end{theorem}

\begin{Proof}
The proof is based on a scheme where we apply the LPIR scheme on the new graph $G=([N]\setminus \mathcal{O}, \mathcal{K})$ when downloading $\theta \in \mathcal{K}$, in addition to downloading the extra messages in $\mathcal{K}'$ if the server is contacted in the LPIR scheme when retrieving $\theta$. In the case when $\theta =i\in \mathcal{K}'$, we identify the unique server, $n\in [N]\setminus \mathcal{O}$, that stores the message $W_i$. From this server, we download $D_{\ell,n}$, i.e., the locally private scheme answer sent in response to $\theta=\ell$ where $\ell\in \mathcal{K}$, alongside the clear messages indexed by $\mathcal{I}_n\cap \mathcal{K}'$. 
\end{Proof}

In the next theorem, we define a notion of modified edge server privacy for general \emph{trees}. In our context, a tree refers to any connected graph which has \emph{leaves}, i.e., nodes whose degree is $1$, and is not necessarily acyclic. Equivalently, we refer to the servers corresponding to the leaves as edge-servers, each of which stores a single message.

\begin{definition}[Modified edge-server privacy]\label{def_modified_edge_main}
    Given a tree graph $G$, let $\mathcal{L}\subset[N]$ denote the set of leaf nodes. Corresponding to a leaf node $l\in \mathcal{L}$, let $\phi(l)$ be its unique parent node. Under local privacy with modified-edge server privacy, the privacy sets are given by
    \begin{align}
        \mathcal{P}_n &=
        \begin{cases}
           \mathcal{I}_{\phi(n)} , & n\in \mathcal{L},\\
           \mathcal{I}_n , & n\in [N]\setminus \mathcal{L}.
        \end{cases}
    \end{align}    
\end{definition}

\begin{remark}
    Note that the mapping $\phi:\mathcal{L}\to [N]\setminus \mathcal{L}$ is surjective, but not injective, since a parent node may have more than one leaf.
\end{remark}

\begin{theorem}\label{thm:tree_modified edge}
    Let $D_k$ be the download cost of a local PIR scheme $\Pi_{LPIR}$, associated with $\theta=k$, and let $D_{k,n}$ denote the corresponding download from server $n$. For a tree graph $G$ under modified edge-server privacy of Definition \ref{def_modified_edge_main}, the capacity is lower bounded as
    \begin{align}
        C\geq \max_{\Pi_{LPIR}} \ \frac{KL}{\sum_{k\in [K]}D_{k}+\sum_{n\in \mathcal{L}} \sum_{k\in \mathcal{P}_n\setminus \{\ell\}:~\mathcal{I}_n=\{\ell\}}(D_{\ell,n}-D_{k,n})}.
    \end{align}
\end{theorem}

\begin{Proof}
The achievability proof builds on any local PIR scheme, and modifies it to suit the modified privacy constraint. Note that, to preserve privacy, the user now downloads more from the servers in $\mathcal{L}$, while the downloads from the remaining servers stay the same. Fix a local PIR scheme $\Pi_{LPIR}$. Let the answer from server $n$ when $\theta=k$ for $k\in \mathcal{P}_n\setminus \mathcal{I}_n$ incur the download cost of $D_{k,n}$. For $\theta=\ell$ where $\mathcal{I}_n=\{\ell\}$, and $n\in\mathcal{L}$, the number of downloaded symbols from server $n$ is $D_{\ell, n}$. In either case, the download is either $\emptyset$ or symbols of $W_{\ell}$ only. For every $n\in \mathcal{L}$, when $k\notin \mathcal{I}_n$, the user queries for the same answers that it downloads from server $n$ if $\theta=\ell$, where $\mathcal{I}_n=\{\ell\}$, which leads to the additional download cost. 
\end{Proof}

\begin{remark}
    If $G$ is the star graph $\mathbf{S}_N$, where server $N$ stores all the messages $\mathcal{W}$ and $\mathcal{W}_n = \{W_n\}$, $n\in [N-1]$, then $\mathcal{L} = [N-1]$ and $\phi(l)= N$ for all $l\in \mathcal{L}$. As a result, the modified edge privacy setting reduces to the GPIR setting.
\end{remark}

\section{Results on Path Graphs}\label{sec_results_path}
For a path graph on $N$ nodes $\mathbf{P}_N$, the storage of server $n$ is  $\mathcal{W}_n  =  \{W_n, W_{n-1}\}$, $n\in [2:N-1]$, with edge-server storage $\mathcal{W}_1 = \{W_1\}$ and $\mathcal{W}_N = \{W_{N-1}\}$. 

\begin{theorem}[Path graphs with local privacy and modified edge-servers privacy] \label{thm:converse_priv_set1_path}
    For the path graph $\mathbf{P}_N$ with modified edge-servers privacy using Definition \ref{def_modified_edge_main}, i.e., the privacy sets are given by $\mathcal{P}_n  = \mathcal{I}_n$, $n \in [2:N-1]$, and $\mathcal{P}_1  = \mathcal{P}_2$ and $\mathcal{P}_N  = \mathcal{P}_{N-1}$, the capacity is  
    \begin{align}
        C = \frac{N-1}{2N-3}.
    \end{align}
\end{theorem}

\begin{remark}
    In comparison with the GPIR for $\mathbf{P}_N$, we note that for $N\geq 4$, the capacity for the path graph with modified edge-servers privacy is higher than the capacity for the aforementioned setting, i.e., $\frac{N-1}{2N-3} > \frac{2}{N}$ \cite{shreya_graph_3}.
\end{remark}

\begin{remark}
    Since $\mathbf{P}_N$ is a tree, from Theorem~\ref{thm:tree_modified edge} the capacity for the modified edge-server setting is upper bounded by the LPIR capacity. In particular, compared to LPIR on $\mathbf{P}_N$, when $N$ is odd\footnote{We compare with the odd number of servers, $N$, because the LPIR capacity is resolved in this case, unlike for the even number of servers case.}, the capacity of LPIR is strictly higher, since $C_{LPIR}(\mathbf{P}_N) = \frac{N-1}{2N-4} > \frac{N-1}{2N-3}$ \cite{local_pir_jour}.
\end{remark}

\begin{Proof}
For the upper bound, for $\theta = k$, we bound the download cost $D_k$ as,
\begin{align}
    D_k &\geq \sum_{n=1}^N H(A_n^{[k]}|\mathcal{Q})\geq H(A_{[N]}^{[k]}|\mathcal{Q})\\
    &=   H(W_k|A_{[N]}^{[k]},\mathcal{Q}) +H(A_{[N]}^{[k]}|\mathcal{Q})\\   
    &= H(W_k|\mathcal{Q})+  H(A_{[N]}^{[k]}|W_k,\mathcal{Q})\\
    & = L+ I(\mathcal{W}\setminus \{W_k\};A_{[N]}^{[k]}|W_k,\mathcal{Q})\label{interference_eq}.
\end{align}
The proof relies on showing the following inequalities,
\begin{align}
    I(\mathcal{W}\setminus\{W_k\};A_{[N]}^{[k]}|W_k, \mathcal{Q})\geq 
    \begin{cases}
        H(A_2^{[2]}|W_1,\mathcal{Q}), & k=1\\
        \frac {L}{2}+H(A_3^{[3]}|W_1,W_2,\mathcal{Q}), & k=2\\
        H(A_k^{[k-1]}|W_k,\mathcal{Q})+H(A_{k+1}^{[k+1]}|\mathcal{W}_{k},\mathcal{Q}), & k\in [3:N-3]\\
        \frac{L}{2}+H(A_{N-2}^{[N-3]}|W_{N-1}, W_{N-2}, \mathcal{Q}), & k=N-2\\
        H(A_{N-1}^{[N-2]}|W_{N-1},\mathcal{Q}), & k=N-1.
    \end{cases} \label{relied-on}
\end{align}
Note that the second and fourth inequalities are introduced in this setting, while the others are the same as those in local PIR \cite{local_pir_conf}. This is because the number of servers that queries for $W_2$ and $W_{N-2}$ require privacy from has increased, compared to that in local PIR. We show the proof of $k=2$, and the case of $k=N-2$ follows similarly,
\begin{align}
    I(\mathcal{W}\setminus \{W_2\};&A_{[N]}^{[2]}|W_2,\mathcal{Q}) \notag\\&\geq I(W_1,W_3;A_1^{[2]},A_2^{[2]},A_3^{[2]}|W_2, \mathcal{Q})\\
    &\geq I(W_1;A_1^{[2]},A_2^{[2]}|W_2, \mathcal{Q}) + I(W_3;A_3^{[2]}|W_1,W_2,\mathcal{Q})\\
    &\geq \frac{1}{2}I(W_1;A_1^{[2]}|W_2,\mathcal{Q})+\frac{1}{2}I(W_1;A_2^{[2]}|W_2,\mathcal{Q}) +  I(W_3;A_3^{[2]}|W_1,W_2,\mathcal{Q}) \label{c_1}\\
    & = \frac{1}{2}H(A_1^{[1]}|W_2,\mathcal{Q})+\frac{1}{2}H(A_2^{[1]}|W_2,\mathcal{Q})+H(A_3^{[3]}|W_1,W_2,\mathcal{Q})\label{c_2}\\
    &\geq \frac{1}{2}H(A_1^{[1]},A_2^{[1]}|W_2,\mathcal{Q}) +\frac{1}{2}H(W_1|A_1^{[1]},A_2^{[1]},W_2,\mathcal{Q}) +H(A_3^{[3]}|W_1,W_2,\mathcal{Q})\\
    &= \frac{L}{2} + H(A_3^{[3]}|W_1,W_2,\mathcal{Q}),
\end{align}
where $\eqref{c_1}$ is due to $2 I(W_1;A_1^{[2]},A_2^{[2]}|W_2,\mathcal{Q}) = I(W_1;A_1^{[2]}|W_2,\mathcal{Q})+ I(W_1;A_2^{[2]}|A_1^{[2]}, W_2,\mathcal{Q})+ I(W_1;A_2^{[2]}|W_2,\mathcal{Q})+ I(W_1;A_1^{[2]}|A_2^{[2]}, W_2,\mathcal{Q})$, and \eqref{c_2} is due to privacy. 

Now, by summing \eqref{interference_eq} over all $k$, and using \eqref{relied-on}, we obtain,
\begin{align}
     \sum_{k=1}^{N-1}  D_k -(N-1)L &\geq  H(A_2^{[2]}|W_1,\mathcal{Q})+ \frac{L}{2} +H(A_3^{[3]}|W_1,W_2,\mathcal{Q})\notag \\
    &\quad +\sum_{k=3}^{N-3} \left( H(A_k^{[k-1]}|W_k,\mathcal{Q}) +H(A_{k+1}^{[k+1]}|W_k,  W_{k-1},\mathcal{Q}) \right)\notag\\
    &\quad+ \frac{L}{2}+H(A_{N-2}^{[N-3]}|W_{N-1},W_{N-2},\mathcal{Q}) +H(A_{N-1}^{[N-2]}|W_{N-1},\mathcal{Q})\\
    &= L+ H(A_2^{[2]}|W_1,\mathcal{Q})+ H(A_3^{[3]}|W_1,W_2,\mathcal{Q})+ H(A_3^{[2]}|W_3, \mathcal{Q}) \notag\\
    &\quad+ H(A_4^{[4]}|W_3,W_2,\mathcal{Q})+\ldots+ H(A_{N-3}^{[N-4]}|W_{N-3},\mathcal{Q}) \nonumber \\ &\quad+H(A_{N-2}^{[N-2]}|W_{N-2},W_{N-3}, \mathcal{Q}) +H(A_{N-1}^{[N-2]}|W_{N-1},\mathcal{Q})\\
    &\geq L+ H(A_2^{[2]},A_3^{[2]}|\mathcal{W}\setminus \{W_2\},\mathcal{Q})+\sum_{k=3}^{N-2} H(A_k^{[k]}, A_{k+1}^{[k]}|\mathcal{W}\setminus \{W_k\},\mathcal{Q})\\
    &= L+ H(W_2,A_2^{[2]},A_3^{[2]}|\mathcal{W}\setminus \{W_2\},\mathcal{Q})\nonumber \\ & \quad +\sum_{k=3}^{N-2} H(W_k,A_k^{[k]}, A_{k+1}^{[k]}|\mathcal{W}\setminus \{W_k\},\mathcal{Q})\\
    &= L+H(W_2|\mathcal{W}\setminus \{W_2\},\mathcal{Q})+H(A_2^{[2]},A_3^{[2]}|\mathcal{W},\mathcal{Q})\notag \\
    &\quad+\left(  \sum_{k=3}^{N-2} H(W_k|\mathcal{W}\setminus\{W_k\},\mathcal{Q})+ H(A_k^{[k]}, A_{k+1}^{[k]}|\mathcal{W},\mathcal{Q})\right)\\
    &= (N-2)L. \label{conv-last-step}
\end{align}
Thus, \eqref{conv-last-step} gives $\sum_{k=1}^{N-1}  D_k -(N-1)L\geq (N-2)L$, hence $\sum_{k=1}^{N-1}  D_k \geq (2N-3)L$. This gives the desired converse,
\begin{align}
    R =\frac{\sum_{k=1}^K L_k}{\sum_{i=k}^K D_k}=\frac{(N-1)L}{\sum_{i=k}^K D_k}\leq \frac{(N-1)L}{(2N-3)L}= \frac{(N-1)}{(2N-3)}.
\end{align}

For achievability, we have two cases. The first case is $\theta \in \{1, N-1\}$ and the second case is $\theta \in [2:N-2]$. For each $k\in [N-1]$, let $L=2$ for all messages, and let $W_k(i)$ denote the $i$th symbol of $W_k$ after permuting the indices of the messages uniformly at random. In the first case, we focus on $\theta =1$ and the other case is handled similarly. In this case, we download $W_1(1)$ from server $1$, $W_1(2)+W_2(1)$ from server $2$ and $W_2(1)$ from server $3$. For the second case, assume without loss of generality that $\theta =2$. Then, we download $W_1(1)$ from server $1$, $W_1(1)+W_2(1)$ from server $2$ and $W_2(2)+W_3(1)$ from server $3$ and $W_3(1)$ from server $4$. The total downloaded symbols is given by $D = 2(3) + (N-3)(4) = 2(2N-3)$ and the rate is given by $R = \frac{2(N-1)}{2(2N-3)} = \frac{N-1}{2N-3}$, concluding the achievability proof.
\end{Proof}

\begin{theorem}[Path graphs with one-sided $h$-neighbor privacy] \label{thm_h_shift_one-sided_path}
     Given the path storage graph $\mathbf{P}_N$ with one-sided $h$-neighbor privacy sets 
     \begin{align}
         \mathcal{P}_n = [\max(1,n-1):\min(n+h,N-1)],
     \end{align}
     the capacity is lower bounded by 
    \begin{align}
        C \geq \frac{2(N-1)}{\frac{(h+2)(h+1)}{2}+3h+5+(h+4)(N-h-3)}, \label{thm6-rate-expression}
    \end{align}
   where $h\in \{1,\ldots,N-2\}$.
\end{theorem}

\begin{remark}\label{rmk_c_lpir lower}
    When $h=0$, this reduces to the local privacy setting, for which, we use the scheme for local PIR in \cite{local_pir_conf,local_pir_jour} where $C_{LPIR}(\mathbf{P}_N)$ is lower bounded by $\frac{N-1}{2N-3}$ when $N$ is even and is equal to $\frac{N-1}{2N-4}$ when $N$ is odd \cite{local_pir_conf,local_pir_jour}. Interestingly, since there is no value of $h$ for which the privacy sets cover all message indices, we do not recover the GPIR setting.
\end{remark}

\begin{remark}
    The main reason why the case $h = 0$ is not considered here is that the scheme for $h=0$ is dependent on the bipartite graph structure of $\mathbf{P}_N$ and requires downloading all messages in $\mathcal{W}_n$ whenever server $n$ is contacted, for privacy. However, for $h>0$, as the privacy sets become larger and non-symmetric, the bipartite scheme becomes less efficient.
\end{remark}

\begin{Proof}
There are three cases in this setting. The first is when $\theta \in [1:h+1]$, in which case, we download $(\theta+2)$ symbols from server $1$ to server $(\theta+2)$ as follows. If $\theta =1$, we download $W_{1}(1)$ from server $1$ and use it as an information symbol. From the second server, we download $W_1(2)+W_2(1)$ and from the third server, we download $W_{2}(1)$. If $\theta\neq 1$, we download $W_1(1)$ to be used as side information. Further, we download $W_{\theta - 1}(1)+W_{\theta}(1)$ from the $\theta$th server and $W_{\theta}(2)+W_{\theta + 1}(1)$ from the $(\theta+1)$th server, $W_{\theta + 1}(1)$ from $(\theta+2)$th server and $W_{k-1}(1)+W_k(1)$ from the $k$th server, where $k \in [2:\theta-1]$. The second case is when $\theta \in [h+2:N-2]$, in which case, we download $h+4$ symbols starting with server $k-1$ when $\theta = h+k $ as a clean symbol for the largest index message, i.e., $W_{k}(1)$, which acts as side information. Similarly, server $k+h+2$ acts as a source of side information, where $W_{k+h+1}(1)$ is downloaded, and for the rest of the servers in between, we download $W_{m-1}(1)+W_{m}(1)$ for $m \in [k:k+h+1] \setminus \{\theta-1,\theta\}$, $W_{\theta-1}(1)+W_{\theta}(1)$ from the $\theta$th server and $W_{\theta}(2)+W_{\theta+1}(1)$ from the $(\theta+1)$th server. For the case when $\theta = N-1$, we download one symbol from the last $h+3$ servers in a similar fashion. Thus, the sum of all downloaded symbols is given by
\begin{align}
    \sum_{\theta = 1}^{N-1} D_{\theta} &= \left(\sum_{\theta=1}^{h+1} \theta+2\right) + \left(\sum_{\theta = h+2}^{N-2} h+4\right) + h+3 \\
    &= \frac{(h+2)(h+1)}{2}+3h+5+(h+4)(N-h-3),
\end{align}
and the rate expression in \eqref{thm6-rate-expression} follows.
\end{Proof}

\begin{theorem}[Path graphs with two-sided $h$-neighbor privacy]\label{thm:h_shift two-sided_rate path_1}
   Given the path storage graph $\mathbf{P}_N$ with two-sided $h$-neighbor privacy sets  
        \begin{align}
            \mathcal{P}_n = [\max(1,n-h-1): \min(n+h,N-1)],
        \end{align}
    the capacity is lower and upper bounded as
    \begin{align}
         \frac{2(N-1)}{(h+2)(2N-h-3)} \leq C \leq 
        \begin{cases}
        \frac{2(N-1)}{(h+2)(2N-h-3)-(h+1)(h+2)}, & 0\leq h \leq \lceil \frac{N-4}{2}\rceil, \\
        \frac{2(N-1)}{(h+2)(2N-h-3)-(N-h-1)(N-h-2)}, &\lceil \frac{N-4}{2}\rceil<h\leq N-3,
        \end{cases} \label{thm7-rate-expression}
    \end{align}
    where $h\in \{1,\ldots,N-3\}$. 
\end{theorem}

\begin{remark}
    When $h=0$, we recover the local privacy setting, and we obtain the same result as the first sentence of Remark~\ref{rmk_c_lpir lower}. On the other hand, when $h=N-2$, each privacy set is $[K]$, yielding $C_{PIR}(\mathbf{P}_N)=\frac{2}{N}$.    
\end{remark}

\begin{Proof}\label{proof_of_thm_3}
We start with the lower bound. For better exposition, we explain the scheme with reference to the GPIR scheme on the path graph given in \cite{shreya_graph_3}. Assume $L_i=2$ for all $i\in [K]$. Let $\theta=k$. The set of servers $\mathcal{N}_k$ that have $k$ in their privacy set is given by
\begin{align}
    \mathcal{N}_k = [\max(1,k-h):\min(N,h+k+1)].
\end{align}
We analyze this in two cases:

\paragraph*{Case 1: $h\leq \lceil\frac{N-4}{2}\rceil$.} In this range, no message index appears in all $\mathcal{P}_n$, $n\in [N]$. This results in the following three ranges of $k$:
\begin{enumerate}
    \item If $k\in [h+1]$, then $\mathcal{N}_k = [1:h+k+1]$. Download one symbol corresponding to the PIR scheme on the path graph consisting of $\mathcal{N}_k$ and server $h+k+2$, when $\theta=k$.
    \item If $k\in [h+2:N-h-2]$, then $\mathcal{N}_k = [k-h:h+k+1]$. Download one symbol corresponding to the PIR scheme on the path graph consisting of $\mathcal{N}_k$, server $k-h-1$ and server $h+k+2$, when $\theta=k$.
    \item If $k\in [N-h-1:N-1]$, then $\mathcal{N}_k = [k-h:N]$. Download one symbol corresponding to the PIR scheme on the path graph consisting of $\mathcal{N}_k$ and server $k-h-1$, when $\theta=k$.
\end{enumerate}
Let $M_k = |\mathcal{N}_k|+1$, $c_{1,k}=h+k+2$, $c_{2,k} = 2h+4$ and $c_{3,k} = N-k+h+2$. Then,
\begin{align}
    R&=\frac{2(N-1)}{\sum_{k=1}^{h+1}M_k+\sum_{k=h+2}^{N-h-2}(M_k+1)+\sum_{k=N-h-1}^{N-1}M_k}\\
    &=\frac{2(N-1)}{\sum_{k=1}^{h+1}c_{1,k}+\sum_{k=h+2}^{N-h-2}c_{2,k}+\sum_{k=N-h-1}^{N-1} c_{3,k}}\\
    &=\frac{N-1}{\sum_{k=1}^{h+1}(h+k+2)+(h+2)(N-2h-3)}\\
    &=\frac{2(N-1)}{(h+2)(2N-h-3)}.
\end{align}

\paragraph*{Case 2: $h>\lceil\frac{N-4}{2} \rceil$.} In this range, there exists at least one $k$ that is present in the privacy set of all servers. This results in the following three ranges of $k$:
\begin{enumerate}
    \item If $k\in [1:N-h-2]$, then $\mathcal{N}_k = [1:h+k+1]$. Download one symbol corresponding to the PIR scheme on the path graph consisting of $\mathcal{N}_k$ and server $h+k+2$, when $\theta=k$.
    \item If $k\in [N-h-1:h+1]$, then $\mathcal{N}_k = [N]$. Download one symbol corresponding to the PIR scheme on path graph $\mathbf{P}_N$, when $\theta=k$.
    \item If $k\in [h+2:N-1]$, then $\mathcal{N}_k = [k-h:N]$. Download one symbol corresponding to the PIR scheme on the path graph consisting of $\mathcal{N}_k$ and server $k-h-1$, when $\theta=k$.
\end{enumerate}
Then,
\begin{align}
    R&=\frac{2(N-1)}{\sum_{k=1}^{N-h-2}M_k+\sum_{k=N-h-1}^{h+1}N+\sum_{k=h+2}^{N-1}M_k}\\
    &=\frac{2(N-1)}{\sum_{k=1}^{N-h-2}c_{1,k}+\sum_{k=N-h-1}^{h+1}N+\sum_{k=h+2}^{N-1} c_{3,k}}\\
    &=\frac{2(N-1)}{2\sum_{k=1}^{N-h-2}(h+k+2)+(2h+3-N)N}\\
    &=\frac{2(N-1)}{(h+2)(2N-h-3)}.
\end{align}
This concludes the proof of the lower bound in \eqref{thm7-rate-expression}. 

For the upper bound, we start by lower bounding the download cost when $\theta=k$ as 
\begin{align}
    D_k\geq L+ I(\mathcal{W}\setminus \{W_k\}; A_{[N]}^{[k]}|W_k,\mathcal{Q}),\label{eq:download_bound}
\end{align}
and derive the following inequalities on the interference $T_k=I(\mathcal{W}\setminus \{W_k\};A_{[N]}^{[k]}|W_k,\mathcal{Q})$ term. We analyze it in the same two ranges of $h$ as the achievable scheme:

\paragraph{Case 1: $h\leq \lceil \frac{N-4}{2}\rceil$.} Then, if $k\in [h+1]$, $k$ is present in the privacy sets of servers $[h+k+1]$. For $k=1$, 
\begin{align}
    T_1 &\geq I(W_{2:h+2};A_{[N]}^{[1]}|W_1,\mathcal{Q})\\
    & = \sum _{l=2}^{h+2} I(W_l;A_{[N]}^{[1]}|W_{1:l-1},\mathcal{Q})\label{eq:T_1 bound start}\\
    &\geq \sum_{l=2}^{h+2} I(W_l;A_l^{[1]}|W_{1:l-1},\mathcal{Q})\\
    &= \sum_{l=2}^{h+2} I(W_l;A_l^{[l]}|W_{1:l-1},\mathcal{Q})\\
    &= \sum_{l=2}^{h+2} H(A_l^{[l]}|W_{1:l-1},\mathcal{Q})\\
    &\geq \sum_{l=2}^{h+2} H(A_l^{[l]}|\mathcal{W}\setminus \{W_l\},\mathcal{Q}),\label{eq:T_1 bound end}
\end{align}
where the last equality holds since $H(A_l^{[l]}|W_{1:l},\mathcal{Q})=0$. 
For $k\in [2:h+1]$, we have
\begin{align}
    T_k&\geq I(W_{1:k-1}, W_{k+1:h+k+1};A_{[N]}^{[k]}|W_k,\mathcal{Q}) \\
    & = I(W_{1:k-1};A_{[N]}^{[k]}|W_k,\mathcal{Q})+I(W_{k+1:h+k+1};A_{[N]}^{[k]}|W_{1:k},\mathcal{Q})\\
    & \geq \sum_{l=2}^{k} I(W_{l-1};A_{l}^{[k]}|W_{l:k},\mathcal{Q})+\sum_{l=k+1}^{h+k+1} I(W_l;A_l^{[k]}|W_{1:l-1}, \mathcal{Q})\\
    & =\sum_{l=2}^{k} I(W_{l-1};A_{l}^{[l-1]}|W_{l:k},\mathcal{Q})+\sum_{l=k+1}^{h+k+1} I(W_l;A_l^{[l]}|W_{1:l-1}, \mathcal{Q})\\
    & =\sum_{l=2}^{k} H(A_{l}^{[l-1]}|W_{l:k},\mathcal{Q})+\sum_{l=k+1}^{h+k+1} h(A_l^{[l]}|W_{1:l-1}, \mathcal{Q})\\
    &\geq\sum_{l=2}^k H(A_l^{[l-1]}|\mathcal{W}\setminus \{W_{l-1}\}) + \sum_{l=k+1}^{h+k+1} H(A_l^{[l]}|\mathcal{W}\setminus \{W_l\}).
\end{align}
Next, for $k\in [h+2:N-h-2]$, $\mathcal{N}_k = [k-h:h+k+1]$, and we similarly obtain
\begin{align}
    T_k\geq  \sum_{l=k-h}^k H(A_l^{[l-1]}|\mathcal{W}\setminus \{W_{l-1}\}) + \sum_{l=k+1}^{h+k+1} H(A_l^{[l]}|\mathcal{W}\setminus \{W_l\}).
\end{align}
Similarly, for $k\in [N-h-1:N-1]$, we have $\mathcal{N}_k'\in [k-h:N]$, therefore for $k\in [k-h:N-2]$,
\begin{align}
    T_k \geq  \sum_{l=k-h}^kH(A_l^{[l-1]}|\mathcal{W}\setminus \{W_{l-1}\})+\sum_{l=k+1}^{N-1} H(A_l^{[l]}|\mathcal{W}\setminus \{W_l\}).
\end{align}
For $k=N-1$, by symmetry with $k=1$, we have
\begin{align}
    T_{N-1}&\geq I(W_{N-2-h:N-2};A_{[N]}^{[N-1]}|W_{N-1}, \mathcal{Q})\\
    &= \sum_{l=2}^{h+2} I(W_{N-l};A_{[N]}^{[N-1]}|W_{N-l+1:N-1},\mathcal{Q})\\
    &\geq \sum_{l=2}^{h+2} I(W_{N-l};A_{N-l+1}^{[N-1]}|W_{N-l+1:N-1},\mathcal{Q})\\
    &= \sum_{l=2}^{h+2} I(W_{N-l};A_{N-l+1}^{[N-l]}|W_{N-l+1:N-1},\mathcal{Q})\\
    &\geq \sum_{l=2}^{h+2} H(A_{N-l+1}^{[N-l]}|\mathcal{W}\setminus \{W_{N-l}\},\mathcal{Q}).
\end{align}
Now, summing over all $k$, and pairing the terms with the same superscript, we obtain
\begin{align}
    \sum_{k=1}^{N-1} T_k &\geq  (h+1)(N-h-3)L.
\end{align}
This yields 
\begin{align}
    \sum_{k=1}^{N-1} D_k &\geq (N-1)L+(h+1)(N-h-3)L\\
    &= \frac{1}{2}(h+2)(2N-h-3)L - \frac{1}{2}(h+1)(h+2)L,
\end{align}
and thus the upper bound in \eqref{thm7-rate-expression} for $h\leq\lceil \frac{N-4}{2}\rceil$ follows.

\paragraph{Case 2: $h>\lceil \frac{N-4}{2}\rceil$.}
For $k\in [1:N-h-2]$, we similarly compute the bounds as
\begin{align}
    T_k\geq \begin{cases}
        \sum_{l=2}^{h+2} H(A_l^{[l]}|\mathcal{W}\setminus \{W_l\}), & k=1,\\
        \sum_{l=2}^k H(A_l^{[l-1]}|\mathcal{W}\setminus \{W_{l-1}\}) + \sum_{l=k+1}^{h+k+1} H(A_l^{[l]}|\mathcal{W}\setminus \{W_l\}),& k\in [2:N-h-2].
    \end{cases}
\end{align}
For $k\in [N-h-1:h+1]$, $\mathcal{N}_k = [N]$. For these indices, the bound in standard GPIR holds, 
\begin{align}
    D_k \geq \frac{N}{2}L=L+\frac{N-2}{2}L,
\end{align}
since the GPIR capacity for path graphs is $\frac{2}{N}$.

For $k\in [h+2:N-1]$, we similarly have
\begin{align}
    T_k\geq \begin{cases}
         \sum_{l=k-h}^k H(A_l^{[l-1]}|\mathcal{W}\setminus \{W_{l-1}\}) + \sum_{l=k+1}^{N-1} H(A_l^{[l]}|\mathcal{W}\setminus \{W_{l}\}),& k\in [h+2:N-2],\\
        \sum_{l=2}^{h+2} H(A_{N-l+1}^{[N-l]}|\mathcal{W}\setminus \{W_{N-l}\},\mathcal{Q}), & k=N-1.
    \end{cases}
\end{align}
Summing across all $k\in [N-1]$, by grouping together answers with the same superscript, we obtain
\begin{align}
    \sum_{k=1}^{N-1}D_k &\geq (N-1)L + \sum_{k=1}^{N-h-2} T_k   + \sum_{k=h+2}^{N-1} T_k+ \sum_{k=N-h-1}^{h+1} \frac{(N-2)}{2}L\\
    &= (N-1)L + h(N-h-2)L+(2h-N+3)\frac{(N-2)L}{2}\\
    &=\frac{(h+2)(2N-h-3)L}{2} - \frac{(N-h-1)(N-h-2)L}{2},
\end{align}
which yields the upper bound for $h>\lceil \frac{N-4}{2}\rceil$ in \eqref{thm7-rate-expression} .
\end{Proof}

\begin{theorem}[Path graphs with two-sided $h$-neighbor and modified edge-servers privacy]\label{thm:modified edge nhbrs}
    Consider the privacy sets for server $n$ in $\mathbf{P}_N$ given by Definition \ref{def_modified_edge_main}, i.e., $\mathcal{P}_1 = \mathcal{P}_2$, $\mathcal{P}_N = \mathcal{P}_{N-1}$, and $\mathcal{P}_n$ for $n\in [2:N-1]$ as
    \begin{align}
        \mathcal{P}_n &= [\max(1,n-h-1): \min(n+h,N-1)],        
    \end{align}
    the capacity is lower and upper bounded as
    \begin{align}
       \frac{2(N-1)}{(h+2)(2N-h-3)} \leq  C \leq \begin{cases}
    \frac{2(N-1)}{(h+2)(2N-h-3)-h(h+1)}, & 0\leq h \leq \lceil \frac{N-6}{2}\rceil,\\
    \frac{2(N-1)}{(h+2)(2N-h-3)-(N-h-4)(N-h-5)}, &\lceil \frac{N-6}{2}\rceil<h\leq N-3,
    \end{cases} \label{thm8-rate-expression}
    \end{align}
     where $h\in \{0,1,\ldots,N-3\}$.
\end{theorem}

\begin{remark}
    When $h=0$, we recover the modified edge-servers privacy setting, as in Theorem~\ref{thm:converse_priv_set1_path}, and the GPIR setting when $h=N-2$. 
\end{remark}

\begin{Proof}
The basics of the scheme is the same as the one in Theorem~\ref{thm:h_shift two-sided_rate path_1}, and the rate is derived similarly, with the following minor changes. The number of servers $\mathcal{N}'_{k}$ that have $k$ in their privacy sets is
\begin{align}
    \mathcal{N}_k' = 
    \begin{cases}
        \mathcal{N}_{h+2} \cup \{1\}, & k=h+2,\\
        \mathcal{N}_{N-h-2} \cup \{N\}, & k = N-h-2,\\
        \mathcal{N}_k, & \text{ otherwise.}
    \end{cases}
\end{align}
We analyze this in two cases:

\paragraph*{Case 1: $h\leq \lceil\frac{N-6}{2}\rceil$.}
In this range, no message index appears in all $\mathcal{P}_n$, $n\in [N]$. This results in the following three ranges of $k$:
\begin{enumerate}
    \item If $k\in [h+2]$, then $\mathcal{N}_k' = [1:h+k+1]$. Download one symbol corresponding to the PIR scheme on the path graph consisting of $\mathcal{N}_k'$ and server $h+k+2$, when $\theta=k$.
    \item If $k\in [h+3:N-h-3]$, then $\mathcal{N}_k' = [k-h:h+k+1]$. Download one symbol corresponding to the PIR scheme on the path graph consisting of $\mathcal{N}_k'$, server $k-h-1$ and server $h+k+2$, when $\theta=k$.
    \item If $k\in [N-h-2:N-1]$, then $\mathcal{N}_k' = [k-h:N]$. Download one symbol corresponding to the PIR scheme on the path graph consisting of $\mathcal{N}_k'$ and server $k-h-1$, when $\theta=k$.
\end{enumerate}
This gives the same rate 
\begin{align}
    R&=\frac{2(N-1)}{2\sum_{k=1}^{h+1}M'_k+\sum_{k=h+3}^{N-h-3}2M'_k+1+\sum_{k=N-h-1}^{N-1}M'_k}\\
    &=\frac{2(N-1)}{\sum_{k=1}^{h+1} M_k+\sum_{k=h+2}^{N-h-2}M_k+1 +\sum_{k=N-h-1}^{N-1}M_k}\\
    &=\frac{2(N-1)}{(h+2)(2N-h-3)},
\end{align}
where $M'_k = |\mathcal{N}'_k|+1$.

\paragraph*{Case 2: $h>\lceil\frac{N-6}{2} \rceil$.} In this range, there exists at least one $k$ that is present in the privacy set of all servers. This results in the following three ranges of $k$:
\begin{enumerate}
    \item If $k\in [1:N-h-3]$, then $\mathcal{N}_k' = [1:h+k+1]$. Download one symbol corresponding to the PIR scheme on the path graph consisting of $\mathcal{N}_k'$ and server $h+k+2$, when $\theta=k$.
    \item If $k\in [N-h-2:h+2]$, then $\mathcal{N}_k' = [N]$. Download one symbol corresponding to the PIR scheme on the path graph $\mathbf{P}_N$, when $\theta=k$.
    \item If $k\in [h+3:N-1]$, then $\mathcal{N}_k' = [k-h:N]$. Download one symbol corresponding to the PIR scheme on the path graph consisting of $\mathcal{N}_k'$ and server $k-h-1$, when $\theta=k$.
\end{enumerate}
\begin{align}
    R&=\frac{2(N-1)}{\sum_{k=1}^{N-h-3}M'_k+\sum_{k=N-h-2}^{h+2}N+\sum_{k=h+3}^{N-1}M'_k}\\
    &=\frac{2(N-1)}{\sum_{k=1}^{N-h-3}c_{1,k}+\sum_{k=N-h-2}^{h+2}N+\sum_{k=h+3}^{N-1} c_{3,k}}\\
    &=\frac{2(N-1)}{(h+2)(2N-h-3)}.
\end{align}
This concludes the derivation of the lower bound in \eqref{thm8-rate-expression}. 

For the upper bound, we proceed by starting from \eqref{eq:download_bound} and deriving the following inequalities on the interference $T_k=I(\mathcal{W}\setminus \{W_k\};A_{[N]}^{[k]}|W_k,\mathcal{Q})$ term. We analyze it in the same two ranges of $h$ as in the achievable scheme:

\paragraph{Case 1: $h\leq \lceil \frac{N-6}{2}\rceil$.} If $k\in [h+2]$, then $k$ is present in the privacy sets of servers $[h+k+1]$. For $k=1$, we obtain the same bound as in the first case of Theorem~\ref{thm:h_shift two-sided_rate path_1}, i.e.,
\begin{align}
    T_1 &\geq \sum_{l=2}^{h+2} H(A_l^{[l]}|\mathcal{W}\setminus \{W_l\},\mathcal{Q}).
\end{align}
For $k=2$, we obtain the bound
\begin{align}
    T_2&\geq I(W_1;A_1^{[2]}, A_2^{[2]}|W_2,\mathcal{Q})+  \sum _{l=3}^{h+3} I(W_l;A_{l}^{[2]}|W_{1:l-1},\mathcal{Q})\\
    &\geq \frac{1}{2}I(W_1;A_1^{[1]}, A_2^{[1]}|W_2,\mathcal{Q}) + \sum _{l=3}^{h+3} I(W_l;A_{l}^{[l]}|W_{1:l-1},\mathcal{Q})\\
    &=\frac{1}{2}H(A_1^{[1]}, A_2^{[1]}|W_2,\mathcal{Q})+\sum _{l=3}^{h+3} H(A_{l}^{[l]}|W_{1:l-1},\mathcal{Q})\\
    &\geq \frac{1}{2}H(A_1^{[1]}, A_2^{[1]}|\mathcal{W}\setminus \{W_1\},\mathcal{Q})+\sum _{l=3}^{h+3} H(A_{l}^{[l]}|\mathcal{W}\setminus W_l,\mathcal{Q})\\
    &=  \frac{L}{2}+\sum _{l=3}^{h+3} H(A_{l}^{[l]}|\mathcal{W}\setminus \{W_l\},\mathcal{Q}).
\end{align}
For $k\in [3:h+2]$, we follow similar steps to obtain
\begin{align}
    T_k\geq \frac{L}{2} + \sum_{l=3}^k H(A_l^{[l-1]}|\mathcal{W}\setminus \{W_{l-1}\}) + \sum_{l=k+1}^{h+k+1} H(A_l^{[l]}|\mathcal{W}\setminus \{W_l\}).
\end{align}
Next, for $k\in [h+3:N-h-3]$, we have
\begin{align}
    T_k\geq  \sum_{l=k-h}^k H(A_l^{[l-1]}|\mathcal{W}\setminus \{W_{l-1}\}) + \sum_{l=k+1}^{h+k+1} H(A_l^{[l]}|\mathcal{W}\setminus \{W_l\}).
\end{align}
Similarly, for $k\in [k-h:N-3]$, we have
\begin{align}
    T_k \geq \frac{L}{2} + \sum_{l=k-h}^kH(A_l^{[l-1]}|\mathcal{W}\setminus \{W_{l-1}\})+\sum_{l=k+1}^{N-2} H(A_l^{[l]}|\mathcal{W}\setminus \{W_l\}).
\end{align}
For $k=N-2$, we have
\begin{align}
    T_{N-2}&\geq I(W_{N-1};A_N^{[N-2]}, A_{N-1}^{[N-2]}|W_{N-2},\mathcal{Q})+  \sum _{l=3}^{h+3} I(W_{N-l};A_{N-l+1}^{[N-2]}|W_{N-1:N-l+1},\mathcal{Q})\\
    &\geq \frac{1}{2}I(W_{N-1};A_N^{[N-1]}, A_{N-1}^{[N-1]}|W_{N-2},\mathcal{Q}) + \sum _{l=3}^{h+3} I(W_{N-l};A_{N-l+1}^{[N-1]}|W_{N-1:N-l+1},\mathcal{Q})\\
    &=\frac{1}{2}H(A_N^{[N-1]}, A_{N-1}^{[N-1]}|W_{N-2},\mathcal{Q})+\sum _{l=3}^{h+3} H(A_{N-l+1}^{[N-l]}|W_{N-1:N-l+1},\mathcal{Q})\\
    &\geq \frac{1}{2}H(A_N^{[N-1]}, A_{N-1}^{[N-1]}|\mathcal{W}\setminus \{W_{N-1}\},\mathcal{Q})+\sum _{l=3}^{h+3} H(A_{N-l+1}^{[N-l]}|\mathcal{W}\setminus \{W_{N-l}\},\mathcal{Q})\\
    &= \frac{L}{2}+\sum _{l=3}^{h+3} H(A_{N-l+1}^{[N-l]}|\mathcal{W}\setminus \{W_{N-l}\},\mathcal{Q}).
\end{align}
For $k=N-1$, the bound is identical to that in the first case of Theorem~\ref{thm:h_shift two-sided_rate path_1}, i.e.,
\begin{align}
    T_{N-1}&\geq \sum_{l=2}^{h+2} H(A_{N-l+1}^{[N-l]}|\mathcal{W}\setminus \{W_{N-l}\},\mathcal{Q}).
\end{align}
Now, summing over all $k$, and pairing the terms with the same superscript, we obtain
\begin{align}
    \sum_{k=1}^{N-1} T_k &\geq (h+1)L + \sum_{l=2}^{h+2} (l-1)L +\sum_{l=h+3}^{N-h-3} (h+1)L + \sum_{l=N-h-2}^{N-2} (N-l-1)L\\
    &= (h+1)(N-h-2)L.
\end{align}
This yields 
\begin{align}
    \sum_{k=1}^{N-1} D_k &\geq (N-1)L+(h+1)(N-h-2)L\\
    &= \frac{1}{2}(h+2)(2N-h-3)L - \frac{1}{2}h(h+1)L,
\end{align}
and the upper bound in \eqref{thm8-rate-expression} for $h\leq\lceil \frac{N-6}{2}\rceil$ follows.

\paragraph{Case 2: $h>\lceil \frac{N-6}{2}\rceil$.}
For $k\in [1:N-h-3]$, we have
\begin{align}
    T_k\geq \begin{cases}
        \sum_{l=2}^{h+2} H(A_l^{[l]}|\mathcal{W}\setminus \{W_l\}), & k=1,\\
        \frac{L}{2} + \sum_{l=3}^{h+3}H(A_l^{[l]}|\mathcal{W}\setminus \{W_l\}),& k=2,\\
        \frac{L}{2} + \sum_{l=3}^k H(A_l^{[l-1]}|\mathcal{W}\setminus \{W_{l-1}\}) + \sum_{l=k+1}^{h+k+1} H(A_l^{[l]}|\mathcal{W}\setminus \{W_l\}),& k\in [3:N-h-3].
    \end{cases}
\end{align}
For $k\in [N-h-2:h+2]$, $\mathcal{N}'_k = [N]$. For these indices, the bound in standard GPIR holds
\begin{align}
    D_k \geq \frac{N}{2}L=L+\frac{N-2}{2}L,
\end{align}
since the GPIR capacity for path graphs is $\frac{2}{N}$.

For $k\in [h+3:N-1]$, we similarly have
\begin{align}
    T_k\geq \begin{cases}
        \frac{L}{2} + \sum_{l=k-h}^k H(A_l^{[l-1]}|\mathcal{W}\setminus \{W_{l-1}\}) + \sum_{l=k+1}^{N-2} H(A_l^{[l]}|\mathcal{W}\setminus \{W_{l}\}),& k\in [h+3:N-3],\\
        \frac{L}{2} + \sum _{l=3}^{h+3} H(A_{N-l+1}^{[N-l]}|\mathcal{W}\setminus \{W_{N-l}\},\mathcal{Q}),& k=N-2,\\
        \sum_{l=2}^{h+2} H(A_{N-l+1}^{[N-l]}|\mathcal{W}\setminus \{W_{N-l}\},\mathcal{Q}), & k=N-1.
    \end{cases}
\end{align}
Summing across all $k\in [N-1]$, by grouping together answers with the same superscript, we obtain
\begin{align}
    \sum_{k=1}^{N-1}D_k &\geq (N-1)L + \sum_{k=1}^{N-h-3} T_k + \sum_{k=N-h-2}^{h+2} \frac{(N-2)}{2}L + \sum_{k=h+3}^{N-1} T_k\\
    &=(h+2)(2N-h-3)L - (N-h-4)(N-h-5)L,
\end{align}
and the upper bound in \eqref{thm8-rate-expression} for $h\geq\lceil \frac{N-6}{2}\rceil$ follows, concluding the proof.
\end{Proof}

\begin{theorem}[Path graphs with local PIR and extremely private messages]\label{thm_path_extreme_private_odd}
    Given a path storage graph $\mathbf{P}_N$ with odd number of servers $N$, let $\mathcal{E} \subset [N-1]$ be the set of message indices that are extremely private, then we have the following three capacity results: \\
    If neither $1$ nor $N-1$ are in $\mathcal{E}$, then
    \begin{align}\label{e_1}
        C = \frac{2(N-1)}{(|\mathcal{E}|+4)N -4|\mathcal{E}|-6}.
    \end{align}
    If exactly one of $1$ or $N-1$ are in $\mathcal{E}$, then
    \begin{align}\label{e_2}
        C= \frac{2(N-1)}{(|\mathcal{E}|+4)N -4|\mathcal{E}|-5}.
    \end{align}
    If both $1$ and $N-1$ are in $\mathcal{E}$, then
        \begin{align}\label{e_3}
        C= \frac{2(N-1)}{(|\mathcal{E}|+4)N -4|\mathcal{E}|-4}.
    \end{align}
\end{theorem}

\begin{Proof}
The achievability proof stems directly from applying the optimal scheme in \cite{local_pir_conf,local_pir_jour} for the messages in $[N-1]\setminus \mathcal{E}$ and the optimal scheme for path graph GPIR in \cite{graphbased_pir} for messages in $\mathcal{E}$. The converse bound is given by Theorem~\ref{general_extreme_privacy}.
\end{Proof}

\begin{remark}
    For the even number of servers case in Theorem~\ref{thm_path_extreme_private_odd}, the rates in \eqref{e_1}, \eqref{e_2} and \eqref{e_3} are achievable. However, they do not yield capacity results since the optimal scheme for the even number of servers case for LPIR is so far unresolved.
\end{remark}

\begin{theorem}[Path graphs with local PIR and oblivious edge-servers]\label{thm_path_obl_1}
    Given the path storage graph $\mathbf{P}_N$ with the first and last servers as oblivious, i.e., $\mathcal{O}=\{1,N\}$, and the remaining servers follow local privacy, i.e., with privacy sets
    \begin{align}
        \mathcal{P}_n = \begin{cases}
            [N-1], & n \in \{1,N\},\\
            \mathcal{I}_n, & n \in \{2,3,\ldots, N-1\},
        \end{cases}
    \end{align}
    the capacity is lower bounded by
    \begin{align}
        C \geq \frac{1}{2},
    \end{align}
    with message lengths $L_1=L_{N-1}=1$ and $L_i = 2$ for $i \in \{2,3, \ldots, N-2\}$.
\end{theorem}

\begin{Proof}
    For any given $\theta$, we download nothing from the two oblivious servers, and download two clear symbols from the servers where $W_{\theta}$ is replicated. Thus, we download $2$ symbols in total if $\theta \in \{1,N-1\}$, and $4$ symbols in total if $\theta\in [2:N-2]$, yielding the rate $\frac{1}{2}$.
\end{Proof}

\begin{remark}
    One can view Theorem~\ref{thm_path_obl_1} as a stricter version of modified edge-servers privacy for path storage graphs compared to Theorem~\ref{thm:converse_priv_set1_path}. In Theorem~\ref{thm_path_obl_1}, we consider them to be completely oblivious, whereas, in Theorem~\ref{thm:converse_priv_set1_path}, we consider them to have the privacy set identical to that of their neighboring servers. Therefore, the capacity result in Theorem~\ref{thm:converse_priv_set1_path}, $\frac{N-1}{2N-3}$ serves as an upper bound for this setting.
\end{remark}

The following theorem is a capacity lower bound on $\mathbf{P}_N$ where for any $\mathcal{O}\in [N]$ such that $[N]\setminus \mathcal{O}$ is a vertex cover of the graph and the remaining servers follow local privacy. 

\begin{theorem}[Path graphs with non-oblivious servers forming a vertex cover]\label{thm_path_obl_2}
   Given the path storage graph $\mathbf{P}_N$, if $\mathcal{O}$ is a subset of the odd-indexed servers and $\mathcal{P}_n=\mathcal{I}_n$ for $n \in [N]\setminus \mathcal{O}$, then, 
    \begin{align}
        C\geq \begin{cases}
            \frac{1}{2}, & N \text{ odd}\\
            \frac{N-1}{2N-3}, & N \text{ even.}
        \end{cases}
    \end{align}
    If $\mathcal{O}$ is a subset of the even-indexed servers, then 
        \begin{align}
        C\geq \begin{cases}
            \frac{N-1}{2N-4}, & N \text{ odd}\\
            \frac{N-1}{2N-3}, & N \text{ even.}
        \end{cases}
    \end{align}
    In both cases, the length of each message is $L=1$.
\end{theorem}

\begin{Proof}
Let $\mathcal{V}\subseteq [N]\setminus \mathcal{O}$ be the minimal vertex cover for $\mathbf{P}_N$. For both cases, the achievability for $\theta=k$ holds by downloading all the messages from the specific server in the set $\mathcal{V}$ that stores $W_{\theta}$, and nothing from the remaining servers.
\end{Proof}

\begin{remark}
It is clear that when $N$ is odd and $\mathcal{O}$ is \emph{any} subset of odd-indexed servers, the rate is the same as in Theorem~\ref{thm_path_obl_1}. Thus, including servers in addition to the edge-servers in $\mathcal{O}$ does not affect the rate, even though the message lengths are different for the two schemes. In contrast, when $N$ is odd and $\mathcal{O}$ is \emph{any} subset of even-indexed servers, the scheme and the rate coincide with the optimal LPIR setting. Similarly, when $N$ is even, the rate coincides with the best-known LPIR rate for both types of choices of $\mathcal{O}$.
\end{remark}

\section{Results on Cyclic Graphs}\label{sec_results_cyclic}
For a cyclic graph on $N$ nodes $\mathbf{C}_N$, the storage of server $n$ is $\mathcal{W}_n = \{W_{n-1}, W_n\}$ with the convention that $W_0 = W_N$.

\begin{corollary}\label{corollary1}
    For any arbitrary privacy setting, the capacity of the cyclic graph with $N=3$ servers is given by
    \begin{align}
        C=\frac{1}{2}.
    \end{align}
\end{corollary}

The proof of Corollary \ref{corollary1} is a direct consequence of Lemma \ref{basic_ub_lb_lemma} where both the local PIR capacity and the GPIR capacity for $\mathbf{C}_3$ are equal to $\frac{1}{2}$. The GPIR scheme can be used without loss of generality for any arbitrary privacy. 

Although the GPIR scheme can be used for $\mathbf{C}_3$ for any arbitrary privacy, it requires the message subpacketization to be $L=8$ symbols. This can be modified based on the specific arbitrary privacy requirement. As an example, in the case that the first server is oblivious and the other two servers require local privacy, the scheme in Table \ref{tab_cyclic_3} shows that subpacketization with $L=2$ symbols per message is possible. 

\begin{table}[h]
    \centering
    \begin{tabular}{|c|c|c|c|}
    \hline
      & server 1  & server 2 & server 3 \\
      \hline
      $\theta=1$& $a_1, c_1$  & $a_2+b_1$ & $b_1$\\
      \hline
      $\theta=2$& $a_1, c_1$  & $a_1+b_1$ & $b_2+c_1$\\
      \hline
      $\theta=3$& $a_1, c_1$  & $b_1$ & $c_2+b_1$\\
       \hline
    \end{tabular}
    \caption{Scheme for $\mathbf{C}_3$, where the first server is oblivious and the remaining two servers require local privacy.}
    \label{tab_cyclic_3}
\end{table}

\begin{theorem}[Cyclic graph with $1$st neighbor privacy]\label{thm:cyclic-1st}
    For the cyclic graph with $N$ nodes $\mathbf{C}_N$ with $\mathcal{P}_n = \mathcal{I}_n \cup \mathcal{I}_{n+1} $, the capacity is lower and upper bounded as
    \begin{align}
       \frac{2}{5} \leq C \leq \frac{1}{2}.
    \end{align}
\end{theorem}

\begin{remark}
    Note that both lower and upper bounds are independent of the number of servers $N$. This is because, the privacy for each index is confined to a fixed number of servers due to the symmetry of cyclic graphs unlike the path graphs. 
\end{remark}

\begin{remark}
    For $N \geq 5$, the lower bound on the capacity with the $1$st neighbor privacy is higher than the capacity of GPIR for the cyclic graph, where $C_{PIR}(\mathbf{C}_N) = \frac{2}{N+1}$ \cite{karim_graph}.
\end{remark}

\begin{Proof}
For the lower bound, we proceed as follows. By the symmetry of $\mathbf{C}_N$, assume that $\theta=1$ without loss of generality. Let $L=2$ symbols for all messages. The user permutes the message indices using permutations chosen uniformly at random. The user downloads $W_1(1)+W_N(1)$ from server $1$ and $W_1(2)+W_2(1)$ from server $2$. To decode the two message symbols, the user downloads $W_{2}(1)$ from server $3$, $W_N(1)+W_{N-1}(1)$ from server $N$ and $W_{N-1}(1)$ from server $N-1$. Thus, $5$ symbols are downloaded to decode the required two message symbols.

For the upper bound, we have $C_{LPIR}(\mathbf{C}_N)=\frac{1}{2}$. Since the local PIR capacity is always higher than the arbitrary private set capacity, we have the upper bound.
\end{Proof}

\begin{theorem}[Cyclic graph with $i$th neighbor privacy]\label{thm:cyclic_k_k+i}
    For the cyclic graph with $N$ nodes $\mathbf{C}_N$ and $\mathcal{P}_n = \mathcal{I}_n \cup \mathcal{I}_{n+i} $, $i \in [2:N-2]$, the capacity is bounded by,
    \begin{align}
       \frac{1}{3} \leq C \leq \frac{2}{5} .
    \end{align}
\end{theorem}

\begin{remark}
    We have $\frac{2}{5}$ here as an upper bound as opposed to a lower bound for Theorem~\ref{thm:cyclic-1st}. This is because the number of servers that have $k \in \mathcal{P}_n$ has increased for all $k \in [K]$, requiring stricter privacy.
\end{remark}

\begin{Proof}\label{pf_thm_5}
To prove the lower bound, we proceed as follows. Without loss of generality, let $\theta = 1$. From the first server, we download $W_1(1)+W_N(1)$. From the second server, we download $W_1(2)+W_2(1)$. From the third server, we download $W_2(1)$. To cancel the interference here, we have two cases, $i=2$ and $i > 2$. If $i=2$, we download $W_N(1)+W_{N-1}(1)$ from server $N$, $W_{N-1}(1)+W_{N-2}(1)$ from server $N-1$ and $W_{N-2}(1)$ from server $N-2$. If $i > 2$, we download $W_{N}(1)$ from server $N$ and download any linear combination from any two servers other than the first and the second servers, that have $1 \in \mathcal{P}_n$. In all cases, we download $6$ symbols to decode the $2$ message symbols for any $\theta\in [N]$, yielding the rate $R = \frac{2N}{6N} = \frac{1}{3}$.

For the upper bound, recall that
\begin{align}
    D_k \geq L + I(\mathcal{W}\setminus W_k; A_{[N]}^{[k]}|W_k,\mathcal{Q}).
\end{align}
Let $\mathcal{W}^* = \{W_k,W_{k-1},W_{k-i+1},W_{k-i+3},W_{k+1}\}$, then,
\begin{align}
    I(&\mathcal{W}\setminus W_k; A_{[N]}^{[k]}|W_k,\mathcal{Q})\nonumber \\ \geq& I(W_{k-i+1},W_{k-1};A_{[N]}^{[k]}|W_k,\mathcal{Q})+ I(W_{k-i+3},W_{k+1};A_{[N]}^{[k]}|W_k,W_{k-1},W_{k-i+1},\mathcal{Q})\notag\\
    & + I(W_{k-i-1},W_{k-i};A_{[N]}^{[k]}|\mathcal{W}^*,\mathcal{Q}) \\
    \geq & I(W_{k-i+1},W_{k-1};A_{k}^{[k]}|W_k,\mathcal{Q}) + I(W_{k-i+3},W_{k+1};A_{k+1}^{[k]}|W_k,W_{k-1},W_{k-i+1},\mathcal{Q}) \nonumber \\
    &+ I(W_{k-i-1},W_{k-i};A_{k-i}^{[k]},A_{k-i+1}^{[k]}|\mathcal{W}^*,\mathcal{Q})\\
    =& H(A_k^{[k-1]}|W_k,\mathcal{Q}) + H(A_{k+1}^{[k+1]}|W_k,W_{k-1},W_{k-i+1,},\mathcal{Q}) \nonumber\\
    &+ I(W_{k-i-1},W_{k-i};A_{k-i}^{[k]},A_{k-i+1}^{[k]}|\mathcal{W}^*,\mathcal{Q}). \label{c_1_1}
\end{align}
Now, we prove that the third term in \eqref{c_1_1} is lower bounded by $\frac{L}{2}$. For convenience, we denote $\mathcal{W}' = \{W_{k-i-1},W_k,W_{k-1},W_{k-i+1},W_{k-i+3},W_{k+1}\}$. Thus,
\begin{align}
     I(W_{k-i-1},&W_{k-i};A_{k-i}^{[k]},A_{k-i+1}^{[k]}|\mathcal{W}' \setminus W_{k-i-1},\mathcal{Q}) \nonumber \\ &\geq I(W_{k-i};A_{k-i}^{[k]},A_{k-i+1}^{[k]}|\mathcal{W}' ,\mathcal{Q})\\
     &\geq \frac{1}{2}I(W_{k-i};A_{k-i}^{[k]}|\mathcal{W}' ,\mathcal{Q})+ \frac{1}{2}I(W_{k-i};A_{k-i+1}^{[k]}|\mathcal{W}' ,\mathcal{Q})\\
     &= \frac{1}{2}I(W_{k-i};A_{k-i}^{[k-i]}|\mathcal{W}' ,\mathcal{Q})+ \frac{1}{2}I(W_{k-i};A_{k-i+1}^{[k-i]}|\mathcal{W}' ,\mathcal{Q})\\
     &= \frac{1}{2}H(A_{k-i}^{[k-i]}|\mathcal{W}' ,\mathcal{Q}) + \frac{1}{2}H(A_{k-i+1}^{[k-i]}|\mathcal{W}' ,\mathcal{Q})\\
     & \geq \frac{1}{2}H(A_{k-i}^{[k-i]}|\mathcal{W}\setminus W_{k-i} ,\mathcal{Q})  + \frac{1}{2}H(A_{k-i+1}^{[k-i]}|\mathcal{W} \setminus W_{k-i} ,\mathcal{Q})\\
     &\geq \frac{1}{2}H(A_{k-i}^{[k-i]},A_{k-i+1}^{[k-i]}|\mathcal{W}\setminus W_{k-i} ,\mathcal{Q})\\
     & = \frac{1}{2}H(W_{k-i},A_{k-i}^{[k-i]},A_{k-i+1}^{[k-i]}|\mathcal{W}\setminus W_{k-i} ,\mathcal{Q})\\
     &=\frac{L}{2}.
\end{align}
Thus, summing over $k \in [N]$, we have
\begin{align}
    \sum_{k=1}^N D_k & \geq \frac{3NL}{2} + \sum_{k=1}^N H(A_k^{[k-1]}|W_k,\mathcal{Q})  + H(A_{k+1}^{[k+1]}|W_k,W_{k-1},W_{k-i+1,},\mathcal{Q})\\
    & \geq \frac{5NL}{2},
\end{align}
where the last inequality follows the same way as in the proof of Theorem~\ref{thm:converse_priv_set1_path}.
\end{Proof}

\begin{theorem}[Cyclic graph with one-sided $h$-neighbor privacy]\label{thm:h_shift one-sided_rate cycle}
    Consider the privacy sets for server $n$ in $\mathbf{C}_N$ given by
    \begin{align}
    \mathcal{P}_n = [n-1:n+h],
    \end{align}
    where $h\in \{0,1,\ldots,N-2\}$. The capacity is lower bounded as
    \begin{align}
        C \geq \frac{2}{h+4}.
    \end{align}
\end{theorem}

\begin{remark}
    The lower bound coincides with the PIR capacity, $\frac{2}{N+1}$ when $h=N-3$, whereas $\mathcal{P}_n = [N]$, $n\in [N]$ when $h=N-2$. This gap suggests the existence of another scheme with rate $R = \frac{2}{h+3}$, which might be the capacity of this setting.
\end{remark}

\begin{Proof}\label{pf_thm_6}
Assume the message length is $L=2$ and $\theta=k$. Let the user independently permute the symbols of each message, uniformly at random, and denote the permuted message as $W_k = [W_k(1), ~W_k(2)]$. The set of servers $\mathcal{N}_k$ that have $k$ in their privacy set is given by
\begin{align}
    \mathcal{N}_k = [k-h:k+1].
\end{align}
From server $\ell \in [k-h:k]$, the user queries the sum $W_{\ell-1}(1)+W_\ell(1)$. From server $k+1$, the user downloads the sum $W_{k}(2)+W_{k+1}(1)$. Now, from server $k-h-1$, the user downloads $W_{k-h-1}(1)$, and from server $k+2$, the user downloads $W_{k+1}(1)$ as interference symbols. These are used to cancel the unwanted message symbols and recover $W_k$. Thus, $(h+2)+2$ symbols are downloaded to recover $2$ symbols of $W_k$, resulting in the rate $R=\frac{2}{h+4}$.
\end{Proof}

\begin{theorem}[Cyclic graph with two-sided $h$-neighbor privacy]\label{thm:h_shift two-sided_rate cycle}
    Consider the privacy sets for server $n$ in $\mathbf{C}_N$ given as
    \begin{align}
    \mathcal{P}_n = [n-h-1:n+h],
    \end{align}
    where $h\in \{0,1,\ldots,\lfloor\frac{N-3}{2}\rfloor\}$. The capacity is 
    \begin{align}
        C = \frac{1}{h+2}.
    \end{align}
\end{theorem}

\begin{remark}
    When $h=0$, the capacity result coincides with that for LPIR since $C_{LPIR}(\mathbf{C}_N) = \frac{1}{2}$ \cite{local_pir_conf}.
\end{remark}

\begin{Proof}
To prove the achievability, let $\theta = 1$. Then, the index needs to remain private from servers $3$ to $2+h$ on one side, and servers $N$ to $N-h+1$ on the other side, in addition to servers $1$ and $2$. From server $1$, we download $W_1(1)+W_N(1)$, and from server 2, we download $W_1(2)+W_2(1)$. From server $k \in [3:2+h]$, we download $W_{k-1}(1)+W_k(1)$, and similar downloads are made from server $k \in [N-h+1:N]$. Then, from server $3+h$, we download $W_{2+h}(1)$ and the same goes for server $N-h$. Thus, the rate is $R = \frac{2N}{2(h+2)N} = \frac{1}{h+2}$. 

To prove the converse, we have, as \eqref{interference_eq}, for any $k\in [N]$, that $D_k \geq L + I(\mathcal{W}\setminus \{W_k\};A_{[N]}^{[k]}|W_k,\mathcal{Q})$. Let $\mathcal{W}'' = \{W_{k-h-1}, \ldots,W_{k-1},W_{k+1},\ldots,W_{k+h+1}\}$, then
\begin{align}
    I(\mathcal{W}&\setminus \{W_k\};A_{[N]}^{[k]}|W_k,\mathcal{Q})\geq I(\mathcal{W}'';A_{[N]}^{[k]}|W_k,\mathcal{Q})\\
    =& \sum_{l=k-h}^{k} I(W_{l-1};A_{[N]}^{[k]}|W_{[l:k]},\mathcal{Q})  +\sum_{l=k+1}^{k+h+1} I(W_{l};A_{[N]}^{[k]}|W_{[k-h-1:l-1]},\mathcal{Q})\\
    \geq& \sum_{l=k-h}^{k} I(W_{l-1};A_l^{[k]}|\mathcal{W}\setminus \{W_{l-1}\},\mathcal{Q})+\sum_{l=k+1}^{k+h+1} I(W_{l};A_l^{[k]}|\mathcal{W}\setminus \{W_l\},\mathcal{Q})\\
    =& \sum_{l=k-h}^{k} I(W_{l-1};A_l^{[l-1]}|\mathcal{W}\setminus \{W_{l-1}\},\mathcal{Q}) +\sum_{l=k+1}^{k+h+1} I(W_{l};A_l^{[l]}|\mathcal{W}\setminus \{W_l\},\mathcal{Q})\\
    =& \sum_{l=k-h}^k H(A_l^{[l-1]}|\mathcal{W}\setminus \{W_{l-1}\},\mathcal{Q}) +\sum_{l=k+1}^{k+h+1} H(A_l^{[l]}|\mathcal{W}\setminus \{W_l\},\mathcal{Q}).
\end{align}
Then, by summing over all $k \in [N]$,
\begin{align}
    \sum_{k=1}^N D_k-NL &\geq \sum_{k=1}^N\sum_{l=k-h}^k H(A_l^{[l-1]}|\mathcal{W}\setminus \{W_{l-1}\},\mathcal{Q}) +\sum_{k=1}^N \sum_{l=k+1}^{k+h+1} H(A_l^{[l]}|\mathcal{W}\setminus \{W_l\},\mathcal{Q})\\\
    &=(h+1)\sum_{l=1}^N H(A_l^{[l]}|\mathcal{W}\setminus \{W_l\},\mathcal{Q}) +(h+1)\sum_{l=1}^NH(A_{l+1}^{[l]} |\mathcal{W}\setminus \{W_{l}\},\mathcal{Q})\\
    &\geq (h+1) \sum_{l=1}^N H(A_l^{[l]}, A_{l+1}^{[l]}|\mathcal{W}\setminus \{W_l\},\mathcal{Q})\\
    &= (h+1)NL,
\end{align}
which gives $\sum_{k=1}^N D_k \geq (h+2)NL$, thus completing the proof.
\end{Proof}

\begin{theorem}[Cyclic graphs with local PIR and extremely private messages]\label{thm:cyclic_ext_priv}
    Given a cyclic storage graph $\mathbf{C}_N$, let $\mathcal{E} \subset [N]$, be the message indices that are extremely private, while the remaining indices follow local privacy. Then, the capacity is given by
    \begin{align}
        C= \frac{2N}{(|\mathcal{E}|+4)N-3|\mathcal{E}|}
    \end{align}
\end{theorem}

\begin{Proof}
Let $|\mathcal{E}|=n$, we choose $L = 4 \binom{N}{2}$ as the number of message symbols, and apply the scheme for cyclic graphs $\mathbf{C}_N$ proposed in \cite{karim_graph}, to all or the pair of servers $k$ and $k+1$, depending on whether the desired message index $k$ is extremely private or not, respectively. For $k \in \mathcal{E}$, we apply the said scheme exactly to all servers. This involves $\frac{2}{N} \binom{N}{2}(N+1)$ downloaded symbols per server. If $k \notin \mathcal{E}$, then we apply the said scheme only to the servers $k$ and $k+1$ and download the useful side information symbols from the servers $k-1$ and $k+2$. Since the scheme in \cite{karim_graph} requires the user to download $N-1 + 2 \left(\binom{N}{2}- (N-1)\right)$ 2-sum symbols from each server, we download the corresponding $2\left(N-1 + 2 \left(\binom{N}{2}- (N-1)\right)\right)$ clear symbols from servers $k-1$ and $k+2$, to remove the interfering message symbols from the downloaded 2-sums. Thus, the total download is 
\begin{align}
    \sum_{k=1}^N D_k = 2n \binom{N}{2}(N+1) + 2 (N-n)\left( N-1 + 2 \left(\binom{N}{2}- (N-1)\right) + \frac{2}{N} \binom{N}{2}(N+1)\right).
\end{align}
Thus, the rate is given by
\begin{align}
    R
    &=
    \frac{4N {N \choose 2}}
    {2n(N+1){N \choose 2}
    +
    2(N-n)\left(
    N-1+2\left({N\choose 2}-(N-1)\right)
    +\frac{2}{N}{N\choose 2}(N+1)
    \right)} \\[0.5em]
    &=
    \frac{4N \cdot \frac{N(N-1)}{2}}
    {2n(N+1)\frac{N(N-1)}{2}
    +
    2(N-n)\left(
    N-1+2\left(\frac{N(N-1)}{2}-(N-1)\right)
    +\frac{2}{N}\frac{N(N-1)}{2}(N+1)
    \right)} \\[0.5em]
    &=
    \frac{2N^2(N-1)}
    {nN(N-1)(N+1)
    +
    2(N-n)\left(
    N-1+(N-1)(N-2)+(N-1)(N+1)
    \right)} \\[0.5em]
    &=
    \frac{2N^2(N-1)}
    {nN(N-1)(N+1)
    +
    2(N-n)\cdot 2N(N-1)} \\[0.5em]
    &=
    \frac{2N^2(N-1)}
    {N(N-1)\left(n(N+1)+4(N-n)\right)} \\[0.5em]
    &=
    \frac{2N^2(N-1)}
    {N(N-1)\left(nN+n+4N-4n\right)} \\[0.5em]
    &=
    \frac{2N^2(N-1)}
    {N(N-1)\left(nN+4N-3n\right)} \\
    &=
    \frac{2N}{nN+4N-3n} = \frac{2N}{(n+4)N-3n}.
\end{align}

To prove the converse bound, we use the converse bounds for cyclic graphs, from \cite{local_pir_jour} and \cite{karim_graph}, in addition to Theorem~\ref{general_extreme_privacy}. From \cite{karim_graph}, we know that 
\begin{align}
    D_k \geq \frac{L}{2}(N+1),
\end{align}
for all $k \in \mathcal{E}$, from the symmetry of the graph and the privacy requirement. In addition, from \cite{local_pir_jour}, we have
\begin{align}
    D_k \geq 2L,
\end{align}
for $k \in [N] \setminus \mathcal{E}$, again from the symmetry of the graph and the privacy requirement.
Then, by adding both together, we have
\begin{align}
    \sum_{k=1}^N D_k &= \sum_{k=1}^n D_k + \sum_{k=n+1}^N D_k\\
    &\geq \frac{nL}{2}(N+1) + 2(N-n)L\\
    &=NL\left(\frac{n}{2}+2\right) - \frac{3}{2}nL,
\end{align}
and the rate is upper bounded by
\begin{align}
    R=\frac{NL}{\sum_k D_k} \leq \frac{2N}{N(n+4)-3n},
\end{align}
completing the capacity proof.
\end{Proof}

\begin{remark}
    When $n=N$ in Theorem~\ref{thm:cyclic_ext_priv}, we recover the GPIR capacity of the cyclic graph \cite{karim_graph}. In addition, when $n=0$, we recover the LPIR capacity of the cyclic graph \cite{local_pir_jour}.
\end{remark}

\begin{theorem}[Cyclic graphs with non-oblivious servers forming a vertex cover]\label{thm_cyclic_obl_1}
  Let $\mathcal{O} \subset [N]$ be the set of oblivious servers in $\mathbf{C}_N$ and the remaining servers $[N] \setminus \mathcal{O}$ follow local privacy. If $[N] \setminus \mathcal{O}$ forms a vertex cover for $\mathbf{C}_N$, the capacity is given by
    \begin{align}
        C=\frac{1}{2},
    \end{align}
\end{theorem}

\begin{Proof}
    Let $\mathcal{V}\subseteq [N]\setminus \mathcal{O}$ be fixed as the minimal vertex cover for $\mathbf{C}_N$. Then, each message is present in at least one server in $\mathcal{V}$. We set the message lengths to be $2$ for the messages that are stored at servers that are both in $\mathcal{V}$, and the message lengths of the remaining messages as $1$. Let $\theta=k$. We have the following two cases. In the first case, \emph{both} the messages stored at the server in $\mathcal{V}$ storing $W_{k}$ are present in exactly one server in $\mathcal{V}$. For all such $k$, we download both the messages from the server in $\mathcal{V}$ that stores $W_k$. In the second case, \emph{one} of the messages stored at the server in $\mathcal{V}$ storing $W_{k}$ is present in both servers in $\mathcal{V}$. This has two sub-cases. First, if $W_k$ is replicated in servers in $\mathcal{V}$, i.e., servers $\{k,k+1\}\subseteq \mathcal{V}$. Then, we download $W_{k-1}$ and $W_{k+1}$, from server $k$ and server $k+1$, respectively, in addition to downloading one distinct symbol of $W_k$ from each of these servers. Second, if only one of the servers storing $W_k$ is in $\mathcal{V}$, we download $W_k$ from this server, and a single symbol of the other message. For each case, we download double the number of desired message symbols, resulting in the rate $\frac{1}{2}$.
    
    The upper bound comes from the one for the LPIR setting, since the privacy constraints are more strict in this setting, compared to LPIR. One can check that the converse proof in \cite{local_pir_jour} holds in this case with a slight modification to account for the unequal message lengths, yielding the same bound $\frac{1}{2}$.
\end{Proof}

\begin{remark}
    Similar to Theorem~\ref{thm_path_obl_2} for $\mathbf{P}_N$, Theorem~\ref{thm_cyclic_obl_1} is a refinement of the general case in Theorem~\ref{general_oblivious}, specific to $\mathbf{C}_N$. Different from Theorem~\ref{general_oblivious}, which may need to download from all servers in $[N]\setminus \mathcal{O}$ that form a vertex cover (depending on the underlying LPIR scheme), Theorem~~\ref{thm_cyclic_obl_1} downloads only from the servers that form a minimal vertex cover. 
\end{remark}

The following corollaries consider the case where $[N]\setminus \mathcal{O}$ is the minimal vertex cover $\mathcal{V}$ for $\mathbf{C}_N$. We only show the achievability, since the upper bound proof is same as that of Theorem~\ref{thm_cyclic_obl_1}.

\begin{corollary}\label{Thm_C_N_oblivious_capacity_non_overlapping}
    Given the cyclic storage graph with $N$ servers, where $N$ is even. Let the privacy sets be defined as follows
    \begin{align}
        \mathcal{P}_n = \begin{cases}
            \mathcal{I}_n, & n \text{ is even},\\
            [N], & n \text{ is odd},
        \end{cases}
    \end{align}
    i.e., all the odd-indexed servers are oblivious. Then, the capacity is given by
    \begin{align}
        C = \frac{1}{2}.
    \end{align}
\end{corollary}

\begin{Proof}
Let $L_i=1$ for all $i\in [N]$. Clearly, $\mathcal{V}$ is the set of all even-indexed servers. The scheme follows the first case of the proof of Theorem~\ref{thm_cyclic_obl_1} for all desired indices. In particular, if $\theta=k$, we download both the messages from the server in $\mathcal{V}$ with $W_k$ in its storage and nothing from any other server.
\end{Proof}

\begin{corollary}\label{C_N_odd_oblivious_thm}
    Given the cyclic storage graph with $N$ servers, where $N$ is odd. Let the privacy sets be defined as follows
    \begin{align}
        \mathcal{P}_n = \begin{cases}
            \mathcal{I}_n, &  n \in \{2,4,6, \ldots, N-1,N\},\\
            [N], & \textit{otherwise},
        \end{cases}
    \end{align}
    i.e., all the odd-indexed servers except the $N$th server are oblivious. Then, the capacity is given by
    \begin{align}
        C=\frac{1}{2}.
    \end{align}
\end{corollary}

\begin{Proof}
Let $L_{N-1}=2$ and $L_i=1$ for all $i\in [N]\setminus \{N-1\}$. Here, $\mathcal{V}=\{2,4,6,\ldots,N-1,N\}$. We follow the same scheme as in Corollary \ref{Thm_C_N_oblivious_capacity_non_overlapping}, except when retrieving $\theta = N-1$, since it is available in servers $N-1$ and $N$, both of which are in $\mathcal{V}$. We download $W_{\theta-1}$, $W_{\theta}(1)$ from the $(N-1)$st server and $W_{\theta}(2)$, $W_{\theta+1}$ from the $N$th server after permuting the symbols of $W_{N-1}$ uniformly at random. Thus, the rate is given by
\begin{align}
    R = \frac{(N-1)+2}{2(N-1)+4} = \frac{1}{2}, 
\end{align}
giving the desired result.
\end{Proof}

\begin{remark}
    Based on Corollary \ref{Thm_C_N_oblivious_capacity_non_overlapping}, we can see that, in this setting, we do not need to store messages in the oblivious servers. These servers are not contacted during the retrieval process, thus, their storage can be completely eliminated, saving storage space for the servers, without affecting the rate.
\end{remark}

In the next theorem, we generalize this result to the case where $[N]\setminus \mathcal{O}$ is not necessarily a vertex cover.

\begin{theorem}[Cyclic graphs with local PIR and oblivious servers]\label{thm_cycl_general_oblivious}
    Given a cyclic storage graph $\mathbf{C}_N$ with $\mathcal{O} \subset [N]$ being the set of oblivious servers. Then the capacity is lower bounded as
    \begin{align}
        C \geq \frac{\sum_{k \in [N]}\left(2\mathbbm{1}(k \in \mathcal{I}_{\mathcal{L}}, k \notin \mathcal{I}_{\mathcal{O}})+2\mathbbm{1}(k \in \mathcal{I}_{\mathcal{L}}, k \in \mathcal{I}_{\mathcal{O}'})+\mathbbm{1}(k \in \mathcal{I}_{\mathcal{L}}, k \in \mathcal{I}_{\mathcal{O}\setminus\mathcal{O}'})+ \mathbbm{1}(k \in \mathcal{I}_{\mathcal{O}}, k \notin \mathcal{I}_{\mathcal{L}})\right)}{2N|\mathcal{O}'|+ 4 \sum_k \mathbbm{1}(k \in \mathcal{I}_{\mathcal{L}}, k \notin \mathcal{I}_{\mathcal{O}})+2 \sum_k \mathbbm{1}(k \in \mathcal{I}_{\mathcal{L}}, k \in \mathcal{I}_{\mathcal{O}'})+2 \sum_k \mathbbm{1}(k \in \mathcal{I}_{\mathcal{L}}, k \in \mathcal{I}_{\mathcal{O}\setminus \mathcal{O}'})},
    \end{align}
    where $\mathcal{L} = [N] \setminus \mathcal{O}$ is the set of servers that follow local privacy, $\mathcal{O}'$ is the minimal non-contiguous subset of $\mathcal{O}$ which makes $\mathcal{V} = \mathcal{L} \cup \mathcal{O}'$ a vertex cover; the message lengths are set as 2 symbols except for the message indices in $\mathcal{I}_{\mathcal{O}\setminus \mathcal{O}'}$ which have message lengths equal to $1$.
\end{theorem}

\begin{Proof}
    If $W_\theta$ is present in two servers, such that both of them are in $\mathcal{L}$, we download $2$ clear symbols from each of the two servers with the required message as one of them. To ensure privacy, we additionally download $2$ symbols from the oblivious servers in $\mathcal{O}'$, one each of the stored messages. Thus, we retrieve $2$ symbols of $W_{\theta}$ by downloading $4+2|\mathcal{O}'|$ symbols. In the case when $W_\theta$ is present in one server in 
    $\mathcal{L}$, and one of the oblivious servers in $\mathcal{O}'$, we download the same way as in the previous setting. Thus, we get 2 symbols of useful information but only with $2+2|\mathcal{O}'|$ symbols downloaded. If $W_\theta$ is in one of the oblivious servers in $\mathcal{O}$ and one of the oblivious servers in $\mathcal{O}\setminus \mathcal{O}'$, we download 2 clear symbols from the server in $\mathcal{O}'$ with one of them being the required message. Thus, the number of downloaded symbols in this case is $2|\mathcal{O}'|$. Finally, if $W_\theta$ is in one of the servers in $\mathcal{L}$, and one of the servers in $\mathcal{O}\setminus\mathcal{O}'$, we download 2 clear symbols from the server in $\mathcal{L}$ storing the message and to ensure privacy, we download two symbols from each server in $\mathcal{O}'$.
\end{Proof}

To demonstrate the statement of Theorem~\ref{thm_cycl_general_oblivious} with a specific setting, we provide the following example for $\mathbf{C}_6$.

\begin{example}\label{example_outside_2}
    Consider $\mathbf{C}_6$. Let the servers $1$, $4$, $5$ and $6$ be oblivious. We choose $\mathcal{O}' = \{1,5\}$. Thus, $L_1=L_2=2$ and $L_i=1$ for $i \in [3:6]$. Table \ref{tab_example_outside_2} shows the retrieval scheme. The rate of the scheme is $\frac{1}{4}$.

    \begin{table}[h]
        \centering
        \begin{tabular}{|c|c|c|c|c|c|c|}
        \hline
        
        & server 1 & server 2 & server 3 & server 4 & server 5 & server 6\\
        \hline
        $\theta=1$  & $W_1(1),W_6$ & $W_1(2),W_2(1)$ &  &  & $W_4,W_5$ & \\
         \hline
        $\theta=2$ & $W_1(1),W_6$ & $W_1(2),W_2(1)$ & $W_2(2),W_3$ &  & $W_4,W_5$ & \\
         \hline
        $\theta=3$& $W_1(1),W_6$ &  & $W_2(2),W_3$ &  & $W_4,W_5$ & \\
        \hline
        $\theta=4$& $W_1(1),W_6$ &  &  &  & $W_4,W_5$ & \\
        \hline
        $\theta=5$& $W_1(1),W_6$ &  &  &  & $W_4,W_5$ & \\
        \hline
        $\theta=6$& $W_1(1),W_6$ &  &  &  & $W_4,W_5$ & \\
        \hline
        \end{tabular}
        \caption{Scheme for $\mathbf{C}_6$ with $\mathcal{O}=\{1,4,5,6\}$ and $\mathcal{O}'=\{1,5\}$ for Example~\ref{example_outside_2}.}
        \label{tab_example_outside_2}
    \end{table}
\end{example}

\section{The Pyramid Storage Graph}\label{sec_pyramid}
In this section, we introduce an interesting graph structure that has never been investigated in the PIR literature.  The graph is a non-uniform hypergraph with $N$ nodes and $K=\lfloor\frac{N}{2}\rfloor+3$ messages, since each hyper-edge consists of unequal number of vertices. In the equivalent PIR system, the replication factor\footnote{For simple graphs, such as path and cycle graphs, the replication factor is $2$, since each edge consists of two distinct vertices.} of messages ranges from $\lceil N/2\rceil$ to $\lceil N/2\rceil+2$. We name it the pyramid storage scheme and define it as follows. 

\begin{definition}\label{def_pyramid}
    The pyramid storage graph is defined as follows for $i \in [N]$,
    \begin{align}
        \mathcal{W}_i = W_{[\max(1,i-\lceil\frac{N}{2}\rceil +1):\min(i+2,\lfloor\frac{N}{2}\rfloor+3)]}.
    \end{align}
    Thus, if the number of servers $N$ is even, then the storage is given by
    \begin{align}
        \mathcal{W}_i = \begin{cases}
            \{W_{1}, W_2, \ldots, W_{i+2}\} \quad &\text{if } 1 \leq i \leq \frac{N}{2},\\
            \{W_{i-\frac{N}{2}+1},W_{i-\frac{N}{2}+2}, \ldots, W_{\frac{N}{2}+3}\} \quad &\text{if } \frac{N}{2} +1 \leq i \leq N,
        \end{cases}
    \end{align}
    and if the number of servers is odd, then the storage is given by
    \begin{align}
        \mathcal{W}_i = \begin{cases}
            \{W_1, \ldots, W_{i+2}\}, \quad &\text{if } 1 \leq i \leq t,\\
            \{ W_1, \ldots, W_{t+3}\} \quad &\text{if } i = t+1,\\
            \{W_{i-t}, \ldots, W_{t+3}\}, \quad &\text{if } t+2 \leq i \leq 2t+1,
        \end{cases}
    \end{align}
    where $N = 2t+1$.   
\end{definition}

\begin{definition}[Replication pattern]
    The replication pattern for the $k$th message $\mathcal{R}_k$ is the set of servers that stores $W_k$, 
    \begin{align}
        \mathcal{R}_k = \{n \in [N]: W_k \in \mathcal{W}_n\}.
    \end{align}
\end{definition}    

\begin{remark}
    The replication pattern for arbitrary $N$ can be given by
    \begin{align}
         \mathcal{R}_k = \begin{cases}
         \{1, 2, \ldots, \lceil \frac{N}{2}\rceil\}, \quad &\text{if } k=1,\\
         \{ 1,2, \ldots, \lceil \frac{N}{2}\rceil+1\} \quad &\text{if } k=2,\\
         \{k-2, k-1, \ldots, \lceil\frac{N}{2}\rceil+k-1\}, \quad &\text{if } 3 \leq k \leq \lfloor \frac{N}{2}\rfloor+1,\\
         \{\lfloor\frac{N}{2}\rfloor,\lfloor\frac{N}{2}\rfloor+1, \ldots, N\} \quad &\text{if } k=\lfloor\frac{N}{2}\rfloor+2,\\
         \{\lfloor\frac{N}{2}\rfloor+1, \lfloor\frac{N}{2}\rfloor+2, \ldots, N\}, \quad &\text{if } k=\lfloor\frac{N}{2}\rfloor+3.
        \end{cases}
    \end{align}
\end{remark}

For this graph, we present a scheme for local privacy that automatically ensures modified edge privacy ($\mathcal{P}_1=\mathcal{P}_2$ and $\mathcal{P}_N = \mathcal{P}_{N-1}$), as outlined in the following theorem. 

\begin{theorem}[Pyramid graph with modified edge-servers privacy]\label{thm_pyramid}
    Given the pyramid storage graph in Definition \ref{def_pyramid} with the privacy sets $\mathcal{P}_1 = \mathcal{P}_2$, $\mathcal{P}_{N} = \mathcal{P}_{N-1}$, and $\mathcal{P}_k = \mathcal{I}_k$, $k \in [2:N-1]$. The message lengths are given by $L_i = |\mathcal{R}_i|$. Then, if $N$ is even, the capacity has lower bounded as 
    \begin{align}
        C\geq 
        \frac{N^2+10N}{2(N^2+10N+8)} = \frac{t^2+5t}{2t^2+10t+4},
    \end{align}
    where $N=2t$. The capacity is lower bounded for odd $N$ as
    \begin{align}
        C \geq \frac{t^2+6t+3}{2t^2+11t+9},
    \end{align}
    where $N=2t+1$.
\end{theorem}

Prior to proving Theorem~\ref{thm_pyramid}, we provide an example for the retrieval scheme for the case where the number of servers $N$ is even. The odd number of servers case follows the same scheme.

\begin{example}[Pyramid storage for $N=6$ servers]\label{ex_pyramid}
    The storage for this case is given by 
    \begin{align}
        \mathcal{W}_1 &= \{W_1,W_2,W_3\} \\
        \mathcal{W}_2 &= \{W_1,W_2,W_3,W_4\} \\
        \mathcal{W}_3 &= \{W_1,W_2,W_3,W_4,W_5\} \\
        \mathcal{W}_4 &= \{W_2,W_3,W_4,W_5,W_6\} \\
        \mathcal{W}_5 &= \{W_3,W_4,W_5,W_6\} \\
        \mathcal{W}_6 &= \{W_4,W_5,W_6\}. 
    \end{align}
    We represent the messages after permuting their symbol indices uniformly at random as $W_1 = [a_1,a_2,a_3]$, $W_2 = [b_1,b_2,b_3,b_4]$, $W_3 = [c_1,c_2,c_3,c_4,c_5]$, $W_4 = [d_1,d_2,d_3,d_4,d_5]$, $W_5 = [e_1,e_2,e_3,e_4]$ and $W_6 = [f_1,f_2,f_3]$. Table \ref{tab_ex_pyramid_even} provides the retrieval scheme. The rate is $R=\frac{6}{13}$.
    
    \begin{table}[h]
        \centering
        \begin{tabular}{|p{1cm}|p{2cm}|p{2cm}|p{2cm}|p{2cm}|p{2cm}|p{2cm}|}
             \hline
             & server 1 & server 2 & server 3 & server 4 & server 5 & server 6\\
             \hline
             $\theta=1$& $a_1,b_1,c_1$ & $a_2+b_1+c_1+d_1$ & $a_3+b_1+c_1+d_1+e_1$ & $d_1,e_1$ &  & \\
             \hline
             $\theta=2$& $a_1,b_1,c_1$ & $b_2+a_1+c_1+d_1$ & $b_3+a_1+c_1+d_1+e_1$ &  $b_4+c_1+d_1+e_1+f_1$& $d_1,e_1,f_1$ & \\
             \hline
             $\theta=3$& $a_1,b_1,c_1$ & $c_2+a_1+b_1+d_1$ &$c_3+a_1+b_1+d_1+e_1$  & $c_4+b_1+d_1+e_1+f_1$ & $c_5+d_1+e_1+f_1$ & $d_1,e_1,f_1$ \\
             \hline
             $\theta=4$& $a_1,b_1,c_1$ & $d_1+a_1+b_1+c_1$ & $d_2+a_1+b_1+c_1+e_1$ & $d_3+b_1+c_1+e_1+f_1$ & $d_4+c_1+e_1+f_1$ & $d_5,e_1,f_1$\\
             \hline
             $\theta=5$&  & $a_1,b_1,c_1$ &  $e_1+a_1+b_1+c_1+d_1$& $e_2+b_1+c_1+d_1+f_1$ & $e_3+c_1+d_1+f_1$ & $d_1,e_4,f_1$\\
             \hline
             $\theta=6$&  &  & $b_1,c_1$ & $f_1+b_1+c_1+d_1+e_1$ & $f_2+c_1+d_1+e_1$ & $f_3,d_1,e_1$\\
             \hline
        \end{tabular}
        \caption{The retrieval scheme for the pyramid storage graph for $N=6$ in Example~\ref{ex_pyramid}.}
        \label{tab_ex_pyramid_even}
    \end{table}
\end{example}

\begin{Proof}
The achievable schemes for the even and the odd number of servers cases are similar, but the calculations are slightly different. 

Prior to delving into the scheme for the even number of servers case, we calculate $\sum L_i$. It is straightforward to check using $\mathcal{R}_k$, that
\begin{align}
    \sum_{k}L_k &= 2\left(\frac{N}{2}\right)+2\left(\frac{N}{2}+1\right) + \left(\frac{N}{2}-1\right)\left(\frac{N}{2}+2\right)\\
    &= \frac{N^2}{4}+\frac{5N}{2}.
\end{align}
We initiate the scheme after permuting the indices of the messages independently and uniformly at random. If $\theta \in [4]$, we follow the following three steps:

\begin{enumerate}
    \item Download 3 clear symbols from the first server, one from each message, i.e., $W_1(1)$,  $W_2(1)$, $W_3(1)$.
    \item If $\theta \in \mathcal{W}_n$, where $\mathcal{W}_n = \{W_{\theta}, W_{i_1}, W_{i_2}, \ldots, W_{i_m}\}$, download the $(m+1)$-sum as follows
    \begin{align}
        W_{\theta}(\sum_{m\in [n]}\mathbbm{1}(\theta \in \mathcal{W}_{m}))+W_{i_1}(1)+W_{i_2}(1)+\ldots+W_{i_m}(1).
    \end{align}
    \item Otherwise, if $\theta \notin \mathcal{W}_n$, then we download $W_{i_j}(1)$ if $W_{i_j} \in \mathcal{W}_k$ and $W_{\theta} \in \mathcal{W}_k$, if not downloaded previously.
\end{enumerate}   

If $N \geq 12$ and $\theta \in [5:\frac{N}{2}-1]$, we only carry out steps 2 and 3 from the previous case. Finally, if $\theta \in [\frac{N}{2}:\frac{N}{2}+3]$, we change step 1 to downloading 3 clear symbols from the last server, one from each message, i.e., $W_{\frac{N}{2}+1}(1),  W_{\frac{N}{2}+2}(1), W_{\frac{N}{2}+3}(1)$ if $\theta =\frac{N}{2}$, $W_{\frac{N}{2}+1}(|\mathcal{R}_{\frac{N}{2}+1}|),  W_{\frac{N}{2}+2}(1), W_{\frac{N}{2}+3}(1)$ if $\theta =\frac{N}{2}+1$, $W_{\frac{N}{2}+1}(1),  W_{\frac{N}{2}+2}(|\mathcal{R}_{\frac{N}{2}+2}|), W_{\frac{N}{2}+3}(1)$ if $\theta =\frac{N}{2}+2$ and $W_{\frac{N}{2}+1}(1),  W_{\frac{N}{2}+2}(1), W_{\frac{N}{2}+3}(|\mathcal{R}_{\frac{N}{2}+3}|)$ if $\theta =\frac{N}{2}+3$. 

The number of downloaded symbols is given by 
\begin{align}
    D_k = \begin{cases}
        N+1, & \theta \in \{1,\frac{N}{2}+3\},\\
        N+3, & \theta \in \{2,\frac{N}{2}+2\},\\
        N+4, & \text{otherwise}.
    \end{cases}
\end{align}
Thus, $\sum_k D_k = \frac{N^2}{2}+5N+4$, which concludes the proof for the even number of servers case.

For the case of the odd number of servers, we follow the same scheme as the even number of servers. Note that the replication pattern for this case is given by
\begin{align}
     \mathcal{R}_k = \begin{cases}
     \{1, 2, \ldots, t+1\},  &\text{if } k=1,\\
     \{ 1,2, \ldots, t+2\}  &\text{if } k=2,\\
     \{k-2, k-1, \ldots, t+k\}, &\text{if } 3 \leq k \leq t+1,\\
     \{t,t+1, \ldots, N\}  &\text{if } k=t+2,\\
     \{t+1,t+2 \ldots, N\},  &\text{if } k=t+3,
     \end{cases}
\end{align}
where $N=2t+1$. The sum of all message lengths is $\sum_{k}L_k = t^2+6t+3$ and $\sum_k D_k = 2t^2+11t+9$, yielding the lower bound and concluding the proof.
\end{Proof}

\begin{remark}
   One can verify that the aforementioned scheme also works for the case with local privacy without any edge-servers privacy modifications. In addition, the scheme can be slightly modified by moving the side information downloaded from some servers to ensure the privacy in the following case $\mathcal{P}_1=\mathcal{P}_3$, $\mathcal{P}_N=\mathcal{P}_{N-2}$, and $\mathcal{P}_n = \mathcal{I}_n$ otherwise. In both cases, the rate is exactly the same.
\end{remark}

\section{Conclusion and Future Work}\label{sec_conc}
In this paper, we formulated the PIR problem with arbitrary privacy patterns for graph-based storage systems. In our problem formulation, for a given graph-replicated storage setting, its local PIR (LPIR) capacity and graph PIR (GPIR) capacity provide trivial upper and lower bounds, respectively, on the capacity for any arbitrary privacy setting. We derived general capacity bounds for all graphs for well-motivated privacy requirements that combine local privacy with stricter privacy requirements. To make the problem formulation tractable, we fixed the storage to simple graphs, specifically the path and the cyclic storage graphs. For these cases, we derived capacity bounds or exact capacity results for privacy sets that can be defined by the message indices contained in the storage of the neighboring servers. 

Specifically, we derived the exact capacity for the modified edge-servers privacy setting for path graphs. Moreover, the two-sided $h$-neighbor privacy setting demonstrates a unifying framework to move between the LPIR and GPIR settings, through the integer $h$. Next, we fine-tuned the general results of privacy settings with extremely private message subsets, and obtained capacity results for path graphs with odd numbers of vertices, and for cyclic graphs. Similarly, we derived capacity lower bounds and exact capacity for settings containing subsets of oblivious servers. Interestingly, for cyclic graphs with certain subsets of oblivious servers, we designed schemes with unequal message lengths such that the capacity remains the same as that of LPIR, even though the privacy settings are stricter. We then introduced the pyramid storage scheme and provided rate lower bounds for three specific privacy patterns.

We believe that our paper serves as an initial stepping stone and opens up many new directions for future work. Beyond the privacy settings that we explored, one may construct other interesting, well-motivated privacy settings that reflect those that are encountered in practice. Although the LPIR capacity forms a trivial upper bound, finding a general framework to derive upper bounds, not necessarily tight, is an important open problem. Also, analyzing the asymptotic capacity as the number of servers $N$ or the number of messages $K$ grows to infinity is an interesting research direction.

\bibliographystyle{unsrt}
\bibliography{references_2}

@article{c_pir,
  title={The capacity of private information retrieval},
  author={H. Sun and S. Jafar},
  journal={IEEE Trans. Info. Theory},
  volume={63},
  number={7},
  pages={4075-4088},
  month={July},
  year={2017}}

@article{csa,
  title={Cross subspace alignment and the asymptotic capacity of {$X$}-secure {$T$}-private information retrieval},
  author={Z. Jia and H. Sun and S. Jafar},
  journal={IEEE Trans. Info. Theory},
  volume={65},
  number={9},
  pages={5783-5798},
  year={2019},
  month={May}}

@article{chor,
  title={Private information retrieval},
  author={B. Chor and E. Kushilevitz and O. Goldreich and M. Sudan},
  journal={Jour. of the ACM},
  volume={45},
  number={6},
  pages={965-981},
  year={1998},
  month={November}}

@article{banawan_pir_mdscoded,
  title={The capacity of private information retrieval from coded databases},
  author={K. Banawan and S. Ulukus},
  journal={IEEE Trans. Info. Theory},
  volume={64},
  number={3},
  pages={1945-1956},
  year={2018},
  month={January}}

@article{byzantine_tpir,
  title={The capacity of private information retrieval from {B}yzantine and colluding databases},
  author={K. Banawan and S. Ulukus},
  journal={IEEE Trans. Info. Theory},
  volume={65},
  number={2},
  pages={1206-1219},
  year={2018},
  month={September}}

@article{semantic_pir,
  title={Semantic private information retrieval},
  author={S. Vithana and K. Banawan and S. Ulukus},
  journal={IEEE Trans. Info. Theory},
  volume={68},
  number={4},
  pages={2635-2652},
  year={2021},
  month={December}}

@article{nan_eaves,
  title={The capacity of symmetric private information retrieval under arbitrary collusion and eavesdropping patterns},
  author={J. Cheng and N. Liu and W. Kang and Y. Li},
  journal={IEEE Trans. Info. Foren. Security},
  volume={17},
  pages={3037-3050},
  year={2022},
  month={August}}

@article{banawan_eaves,
  title={Private information retrieval through wiretap channel {II}: Privacy meets security},
  author={K. Banawan and S. Ulukus},
  journal={IEEE Trans. Info. Theory},
  volume={66},
  number={7},
  pages={4129-4149},
  year={2020},
  month={February}}

@article{ulukusPIRLC,
    author = {Ulukus, S. and Avestimehr, S. and Gastpar, M. and Jafar, S. and Tandon, R. and Tian, C.},
    journal = {IEEE J. Sel. Areas Commun.},
    title = {Private retrieval, computing, and learning: Recent progress and future challenges},
    month = {March},
    year = {2022},
    volume = {40},
    number = {3},
    pages = {729-748}}

@article{C_SETPIR,
    title={The capacity of private information retrieval with eavesdroppers},
    author={Q. Wang and H. Sun and M. Skoglund},
    journal={IEEE Trans. Info. Theory},
    volume={65},
    number={5},
    pages={3198--3214},
    year={2018},
    month={December}}

@ARTICLE{skoglund_mds_spir,
  author={Wang, Q. and Skoglund, M.},
  journal={IEEE Trans. Info. Theory}, 
  title={Symmetric Private Information Retrieval from {MDS} Coded Distributed Storage With Non-Colluding and Colluding Servers}, 
  year={2019},
  volume={65},
  number={8},
  pages={5160-5175},
  month = {March}}

@INPROCEEDINGS{asymmetric,
  author={Nomeir, M. and Vithana, S. and Ulukus, S.},
  booktitle={CISS}, 
  title={Asymmetric {$X$}-Secure {$T$}-Private Information Retrieval: More Databases is Not Always Better}, 
  year={2024},
month={March}}

@INPROCEEDINGS{karim_graph,
  author={Banawan, K. and Ulukus, S.},
  booktitle={IEEE ISIT}, 
  title={Private Information Retrieval from Non-Replicated Databases}, 
  year={2019},
month = {July}}

@article{shreya_graph_3,
      title={New Capacity Bounds for {PIR} on Graph and Multigraph-Based Replicated Storage}, 
      author={X. Kong and S. Meel and T. Maranzatto and I. Tamo and S. Ulukus},
      journal={IEEE Trans. Info. Theory},
      volume={72},
      number={1},
      pages={691-709},
      month={January},
      year={2026}
}

@ARTICLE{sadeh_graphs,
  author={Sadeh, B. and Gu, Y. and Tamo, I.},
  journal={IEEE Trans. Info. Foren. Security}, 
  title={Bounds on the Capacity of Private Information Retrieval Over Graphs}, 
  year={2023},
  month={November},
  volume={18},
  number={},
  pages={261--273},
  }

@article{shreya_graph_5,
      title={The Role of Common Randomness Replication in Symmetric {PIR} on Graph-Based Replicated Systems}, 
      author={S. Meel and S. Ulukus},
      journal={IEEE Trans. Inf. Theory},
      note={To appear, also available online at arXiv:2602.16700}
}

@ARTICLE{jafar_asymptotic_graph,
  author={Jia, Z. and Jafar, S.},
  journal={IEEE Trans. Info. Theory}, 
  title={On the Asymptotic Capacity of {X}-Secure {T}-Private Information Retrieval With Graph-Based Replicated Storage}, 
  year={2020},
  volume={66},
  number={10},
  pages={6280-6296},
month={July}}

@article{graphbased_pir,
    title={Private information retrieval in graph-based replication systems},
    author={N. Raviv and I. Tamo and E. Yaakobi},
    journal={IEEE Trans. Info. Theory},
    volume={66},
    number={6},
    pages={3590-3602},
    year={2019},
    month={November}}

@inproceedings{nomeir_asymp_bspir,
  title={The Asymptotic Capacity of {B}yzantine Symmetric Private Information Retrieval and Its Consequences},
  author={Nomeir, M. and Aytekin, A. and Ulukus, S.},
  booktitle = {IEEE ISIT},
  year = {2025},
  month = {June}
}

@inproceedings{local_pir_conf,
    title={Local Private Information Retrieval: A New Privacy Perspective for Graph-Based Replicated Systems},
    author={S. Meel and M. Nomeir and S. Ulukus},
    booktitle = {IEEE ITW},
    year = {2026},
    note={To appear, also Available online at arXiv:2605.10872}}

@misc{local_pir_jour,
      title={Local Private Information Retrieval for Graph-Based Replicated Systems}, 
      author={S. Meel and M. Nomeir and S. Ulukus},
      year={2026},
      note={Available online at ar{X}iv:2608.31150}
}

@inproceedings{semantic_tpir,
    author = {M. Nomeir and A. Aytekin and S. Ulukus},
    title = {The Capacity of Semantic Private Information Retrieval with Colluding Servers},
    booktitle = {IEEE Globecom},
    year = {2025},
    month = {December}
}

@Article{mds_breaking,
AUTHOR = {Sun, H. and Tian, C.},
TITLE = {Breaking the {MDS}-{PIR} Capacity Barrier via Joint Storage Coding},
JOURNAL = {Information},
VOLUME = {10},
month={August},
YEAR = {2019},
NUMBER = {9},
ARTICLE-NUMBER = {265},
pages = {2078-2489},
}

@ARTICLE{qpir,
    author={S. Song and M. Hayashi},
    journal={IEEE Trans. Info. Theory}, 
    title={Capacity of Quantum Private Information Retrieval With Multiple Servers}, 
    year={2021},
    month={September},
    volume={67},
    number={1},
    pages={452-463}}

@inproceedings{our_quantum_first,
    title={Quantum Symmetric Private Information Retrieval with Secure Storage and Eavesdroppers}, 
    author={A. Aytekin and M. Nomeir and S. Vithana and S. Ulukus},
    booktitle={IEEE Globecom},
    month={December},
    year={2023}}

@ARTICLE{lu_jafar_byzantine_quantum,
  author={Lu, Y. and Jafar, S.},
  journal={IEEE J. Sel. Areas Info. Theory}, 
  title={Quantum {X}-Secure {T}-Private Information Retrieval From {MDS} Coded Storage With Unresponsive and {B}yzantine Servers}, 
  year={2025},
  volume={6},
  number={},
  pages={59-73}}

@ARTICLE{our_symm_byz_quantum_journal,
    author={M. Nomeir and A. Aytekin and S. Ulukus},
    journal={IEEE Trans. Info. Theory}, 
    title={{B}yzantine-Eavesdropper Alliance: How to Achieve Symmetric Privacy in Quantum {$X$}-Secure {$B$}-{B}yzantine {$E$}-Eavesdropped {$U$}-Unresponsive {$T$}-Colluding {PIR}?}, 
    year={2025}}

@article{LKRAY21_TIT,
  title={Multi-server weakly-private information retrieval},
  author={Lin, H. and Kumar, S. and Rosnes, E. and i Amat, A.  and Yaakobi, E.},
  journal={IEEE Trans. Info. Theory},
  volume={68},
  number={2},
  pages={1197--1219},
  year={2021}
}

@inproceedings{QZTL22,
  title={Improved weakly private information retrieval codes},
  author={Qian, C. and Zhou, R. and Tian, C. and Liu, T.},
  booktitle={IEEE ISIT},
  pages={2827--2832},
  year={2022}
}

@misc{krishnan_graph,
  title={Private Information Retrieval for Graph-based Replication with Minimal Subpacketization},
  author={Shanbhag, V. and Krishnan, P.},
  note={Available online at arXiv:2601.09957},
  year={2026}
}

@INPROCEEDINGS{cyclic_optimal_length,
  author={Keramaati, S. and Salehkalaibar, S.},
  booktitle={IWCIT}, 
  title={Private Information Retrieval from Non-Replicated Databases with Optimal Message Size}, 
  year={2020},
  pages={1-6}}

@misc{gePIR,
  title={Private Information Retrieval over Graphs},
  author={Ge, G. and Wang, H. and Xu, Z. and Zhang, Y.},
  note={Available online at arXiv:2509.26512},
  year={2025}
}

@INPROCEEDINGS{shreya_common_randomness_gspir,
  author={Meel, S. and Ulukus, S.},
  booktitle={IEEE ICC}, 
  title={Effect of Full Common Randomness Replication in Symmetric {PIR} on Graph-Based Replicated Systems}, 
  year={2026},
  month={May}}
\end{document}